\documentclass[11pt,titlepage,a4paper]{article}
\usepackage{bozhomac}  
\usepackage{bozhlogo}  
\usepackage{cite}	
\usepackage{varioref}	

\title{\bfseries	\vspace*{-1.678902345in}
{\huge Pictures and equations of motion\\[1ex]
			in Lagrangian quantum field theory}
}

\vspace{1.7ex}

\author{
Bozhidar Z.\ Iliev
\thanks{Laboratory of Mathematical Modeling in Physics,
Institute for Nuclear Research and \mbox{Nuclear} Energy,
Bulgarian Academy of Sciences,
Boul.\ Tzarigradsko chauss\'ee~72, 1784 Sofia, Bulgaria}
\thanks{E-mail address: bozho@inrne.bas.bg}
\thanks{URL: http://theo.inrne.bas.bg/$^\sim$bozho/}
}

\date{	
 \vspace{2.27ex}\ShortTitle{Picture of motion in QFT}\\[0.27ex]
 \vspace{3.27ex}
\small
	\begin{tabular}{r@{$\colon\to~$}l}
 \vspace{0.09ex} Basic ideas	& May--June, 2001	\\[0.09ex]
 \vspace{0.09ex} Began		& May 12, 2001		\\[0.09ex]
 \vspace{0.09ex} Ended		& August 5, 2001	\\[0.09ex]
 \vspace{0.09ex} Initial typeset& May 18 -- August 8, 2001	\\[0.09ex]
 \vspace{0.09ex} Last update	& January 30, 2003\\[0.09ex]
 \vspace{0.27ex} Produced	& \fbox{\today}	\\[0.27ex]
	\end{tabular} \\[1.27ex]
\normalsize
	\begin{tabular}{r@{$\colon~$}l}
\vspace{0.27ex} http://www.arXiv.org e-Print archive No. & hep-th/0302002
	\end{tabular} \\[-0.27ex]
 \vspace{4.27ex}{\Huge\BOZHO}	\\[4.27ex]
 \vspace{0.27ex}\Subject{Quantum field theory}
								\\[2.27ex]
	\begin{tabular}{r@{\hspace{0.512em}}|@{\hspace{0.512em}}l}
 \vspace{0.27ex}\MSC[2000]{81Q99, 81T99\\\hspace{0pt}}
&
 \vspace{0.27ex}
	\end{tabular} \\[1.27ex]
 \vspace{0.27ex}\\[0.27ex]
}

\newcommand{\Hil}{\mathcal{F}}		
	\newcommand{\base}{\mathit{M}}	

\newcommand{\ope}[2][{}]{\lindex[\mathcal{#2}]{}{#1}} 
\newcommand{\tope}[2][{}]{\ope[#1]{\Tilde{#2}}} 
\begin{document}		

\renewcommand{\thepage}{\roman{page}}

\renewcommand{\thefootnote}{\fnsymbol{footnote}} 
\maketitle				
\renewcommand{\thefootnote}{\arabic{footnote}}   

\tableofcontents		


\begin{abstract}

	The Heisenberg, interaction, and Schr\"odinger pictures of motion are
considered in Lagrangian (canonical) quantum field theory. The equations of
motion (for state vectors and field operators) are derived for arbitrary
Lagrangians which are polynomial or convergent power series in field
operators and their first derivatives. The general links between different
time\ndash dependent pictures of motion are derived. It is pointed that all
of them admit covariant formulation, similar to the one of interaction
picture. A new picture, called the \emph{momentum picture}, is proposed. It
is a 4\ndash dimensional analogue of the Schr\"odinger picture of quantum
mechanics as in it the state vectors are spacetime\ndash dependent, while the
field operators are constant relative to the spacetime. The equations of
motion in momentum picture are derived and partially discussed. In
particular, the ones for the field operators turn to be of algebraic type.
	The general idea of covariant pictures of motion is presented. The
equations of motion in these pictures are derived.

\end{abstract}

\renewcommand{\thepage}{\arabic{page}}

\section {Introduction}
\label{Introduction}

	The aim of the present work is a systematic and detailed theory of
different pictures of motion in Lagrangian quantum field theory and the
derivation of the Euler\ndash Lagrange and Heisenberg equations of motion in
them for general Lagrangians, with or without derivative coupling.

	We should mention, in this paper it is considered only the Lagrangian
(canonical) quantum field theory in which the quantum fields are represented
as operators, called field operators, acting on some Hilbert space, which in
general is unknown if interacting fields are studied. These operators are
supposed to satisfy some equations of motion, from them are constructed
conserved quantities satisfying conservation laws, etc. From the view\ndash
point of present\ndash day quantum field theory, this approach is only a
preliminary stage for more or less rigorous formulation of the theory in
which the fields are represented via operator\ndash valued distributions, a
fact required even for description of free fields. Moreover, in non\ndash
perturbative directions, like constructive and conformal field theories, the
main objects are the vacuum mean (expectation) values of the fields and from
these are reconstructed the Hilbert space of states and the acting in it
fields. Regardless of these facts, the Lagrangian (canonical) quantum field
theory is an inherent component of the most of the ways of presentation of
quantum field theory adopted explicitly or implicitly in books
like~\cite{Bogolyubov&Shirkov,Bjorken&Drell,Roman-QFT,Ryder-QFT,
Akhiezer&Berestetskii,Ramond-FT,Bogolyubov&et_al.-AxQFT,Bogolyubov&et_al.-QFT}.
Besides the Lagrangian approach is a source of many ideas for other
directions of research, like the axiomatic quantum field
theory~\cite{Roman-QFT,Bogolyubov&et_al.-AxQFT,Bogolyubov&et_al.-QFT}.
	By these reasons, we devote the present paper to a general study of
the pictures of motion of Lagrangian field theory.

	The idea for transition from one picture of motion to other one in
quantum field theory is quite simple: it consist in  a simultaneous change
of all state vectors and all operators, in particular the field ones, by
means of a unitary operator in such a way that the mean values (mathematical
expectations) of the operators to remain unchanged.%
\footnote{~%
For a summary and realization of this idea in the Hilbert space and Hilbert
bundle quantum mechanics, see~\cite{bp-BQM-pictures+integrals}; see also the
references cited therein and~\cite[chapter~VII]{Bogolyubov&Shirkov}.%
}
	To avoid confusions, in this paper we shall label all quantities in
the Heisenberg picture by putting tildes (waves) over the
corresponding symbols; \eg $\tope{A}$, $\tope{P}_\mu$, $\tope{\varphi}_i(x)$,
etc. Let $\tope{X}$ be a state vector of a system of quantum fields
$\tope{\varphi}_i(x)$, $i=1,\dots,n\in\field[N]$,
acting on system's Hilbert space $\Hil$ of state
vectors
\footnote{~%
Rigorously speaking, the quantum fields should be regarded as operator\ndash
valued distributions acting on relevant space of test functions in the
correct mathematical setting of quantum field
theory~\cite{Bogolyubov&et_al.-AxQFT,Bogolyubov&et_al.-QFT}. This approach
will be considered elsewhere. Here we treat the quantum fields as operators
acting on a Hilbert space as it is done, e.g.,
in~\cite{Bjorken&Drell-2,Bogolyubov&Shirkov,Itzykson&Zuber}.%
},
and $\tope{A}(x)\colon\Hil\to\Hil$ be an operator
characterizing the system, \eg
$\tope{A}(x)=\tope{\varphi}_i(x),\tope{P}_\mu$, $\tope{P}_\mu$ being the
(canonical, physical) momentum operator of the system defined by its
Lagrangian via Noether's theorem.%
\footnote{~%
For a discussion of the momentum operator in quantum field theory,
see~\cite{bp-QFT-momentum-operator}.%
}
Here and henceforth in this paper by $x$, $x_0$, etc.\ will be denoted points
in the 4\ndash dimensional Minkowski spacetime $\base$, which will serve as
our model of spacetime. If
$\ope{V}(x,x_0)\colon\Hil\to\Hil$, $x,x_0\in\base$, is a unitary operator
relative to the (Hermitian) scalar product
$\clangle\cdot|\cdot\crangle\colon\Hil\times\Hil\to\field[C]$  on $\Hil$,
the changes
	\begin{equation}	\label{12.14}
	\begin{split}
\tope{X} \mapsto \ope{X}^{\ope{V}}
  & := \ope{V}(x,x_0) (\tope{X})
\\
\tope{A}(x) \mapsto \ope{A}^{\ope{V}}(x)
  & := \ope{V}(x,x_0)\circ\tope{X}\circ\ope{V}^{-1}(x,x_0)
	\end{split}
	\end{equation}
preserves the scalar products and mean values, \ie
	\begin{align}	\label{12.15}
\clangle \ope{X}^{\ope{V}} | \ope{Y}^{\ope{V}} \crangle
=
\clangle \tope{X} | \tope{Y} \crangle
\quad
\clangle \ope{X}^{\ope{V}} |
			\ope{A}^{\ope{V}}(x) (\ope{Y}^{\ope{V}}) \crangle
=
\clangle \tope{X} | \tope{A}(\tope{Y}) \crangle .
	\end{align}
Therefore~\eref{12.14} can serve as a transformation from Heisenberg picture,
corresponding to $\ope{V}(x,x_0)=\id_\Hil$,
to the `general $\ope{V}$\ndash picture'.

	In the literature, available to the author, only the case when
$\ope{V}(x,x_0)$ depends solely on the time coordinates $x^0=ct$ and
$x_0^0=ct_0$ of the points $x$ and $x_0$, respectively, has been
investigated; in this case, for brevity, we write $\ope{V}(t,t_0)$ for
$\ope{V}(x,x_0)$. Such pictures and transitions from one picture to other
picture of motion is suitable to be called \emph{time\ndash dependent}
because of their explicit time dependence. However, the example with the
interaction picture, considered in sections~\ref{Sect3} and~\ref{Sect4},
shows that an explicitly time\ndash dependent picture may turn to be
implicitly covariant and, respectively, it may have an explicitly time\ndash
independent formulation (achieved at a price of involving new, more powerful
and difficult, mathematical tools).

	The organization of the present investigation is as follows.

	In section~\ref{Sect2}, the notion of Heisenberg picture of motion in
quantum field theory is briefly recalled and an idea is given of how the
interaction between quantum fields is described in this picture.

	Sections~\ref{Sect3} and~\ref{Sect4} are devoted to the two versions
of the interaction picture, the covariant and time\ndash dependent ones. All
considerations are done in the general case of arbitrary Lagrangian depending
on the quantum fields and their first derivatives. In particular, the
Euler\ndash Lagrange and Heisenberg equations of motion are derived for
arbitrary, derivative or nonderivative, coupling between the fields. Some
non\ndash correct assertions in the literature are corrected. The
calculations and derivations are relatively detailed; one of the reasons
being that they or part of them are directly or, possibly, \emph{mutatis
mutandis} used in the next sections.

	Section~\ref{Sect5} deals with the Schr\"odinger picture of
motion. Regardless of the existence in the literature of some general remarks
concerning this picture, the detailed and  systematic presentation of
Schr\"odinger picture seems to appear in this paper for the first time.
Emphasis is paid to the fact that the Schr\"odinger picture admits a
covariant formulation, similar to the one of interaction picture.

	The idea of a `general' time\ndash dependent picture of motion is
given in section~\ref{Sect6}. The links between arbitrary such pictures are
derived and the equations of motion in them are established. A way is pointed
how all of them can be formulated in a covariant form, similar to that of
interaction picture, based on the Tomonaga\ndash Schwinger equation in
functional derivatives.

	In section~\ref{Sect7} is presented a non-trivial essential
example of completely covariant picture of motion,%
\footnote{~%
We exclude the Heisenberg picture which, by definition, is a covariant one.%
}
called the \emph{momentum picture} as it is completely determined by the
(canonical) momentum operator and has a lot of common features with the
momentum representation (via the Fourier transform) widely applied in quantum
field theory.%
\footnote{~%
However, the momentum picture is \emph{different} from the known momentum
representation. The interrelations between them will be studied elsewhere.%
}
	This new picture is similar to the Schr\"odinger one, the latter
may be considered as its one\ndash dimensional special case. In particular,
in momentum picture the state vectors became spacetime dependent, while the
field operators  (and the observables constructed from them) transform into
constant (in spacetime) operators. Correspondingly, the Euler\ndash Lagrange
field equations transform from second order differential equations in
Heisenberg picture into \emph{algebraic} equations in momentum picture.

	Section~\ref{Sect8} is devoted to some ideas concerning the general
covariant pictures of motion and the equations of motion in them.

	In section~\ref{Conclusion} is summarized the content of the paper.
\vspace{1ex}

	Here are some standard notations and conventions we shall follow in
the present work. The 4\ndash dimensional Minkowski spacetime (model) will be
denoted by $\base$. It is supposed to be endowed with diagonal Lorentz metric
with metric tensor $\eta^{\mu\nu}$ such that
$[\eta^{\mu\nu}]=\diag(+1,-1,-1,-1)$. Here and henceforth in the work, the
Greek indices $\mu,\nu\dots$ run from 0 to $3=\dim\base-1$ and refer to
spacetime coordinates. The raising or lowering of indices is done by
$\eta^{\mu\nu}$ or its inverse tensor $\eta_{\mu\nu}$. The quantum fields
(quantum field operators) are denoted by $\tope{\varphi}_i(x)$ and are
numbered by Latin indices $i,j.\dots$ which run from 1 to some positive
integer $n$. The Einstein's summation convention is accepted, \ie a
summation is understood over any index appearing (twice) at different levels
over the whole range of its values. The coordinates of a point $x\in\base$
are denoted by $x^\mu$ and $\frac{\pd}{\pd x^\mu}$ denotes the partial
derivative with respect to $x^\mu$. If $f$ is a function or operator\ndash
valued function over $\base$, the symbol $\pd_\mu$ denotes an operator such
that
\(
\pd_\mu\colon f\mapsto\frac{\pd f}{\pd x^\mu}
\equiv
\frac{\pd f(x)}{\pd x^\mu}.
\)
By definition $\pd^\mu:=\eta^{\mu\nu}\pd_\nu$.
The composition (product) of mappings/operators will be denoted by $\circ$.
By definition $[A,B]_{\pm}:=A\circ B \pm B\circ A$ for mappings $A$ and $B$
with common domain. The velocity of light in vacuum and Planck's constant
(divided by $2\pi$) are denoted by $c$ and $\hbar$, respectively.

	Ending this section, we would like to make a technical remark. In
the present paper derivatives with respect to operator\ndash valued
(non\ndash commuting) variables will appear often. An everywhere (silently)
accepted procedure for their calculation is by following the rules of
classical analysis of commuting variables with preservation of operator
ordering~\cite{Bjorken&Drell-2,Bogolyubov&Shirkov,Roman-QFT}. However, as it
is demonstrated in~\cite{bp-QFT-action-principle}, that procedure is not
quite correct, but it is harmless in a lot of cases (in particular for free
fields). In~\cite{bp-QFT-action-principle}  is shown that derivatives of
mentioned kind are mappings on the space of initial operators rather than an
operator in this space (as accepted usually). In the sense clarified
in~\cite{bp-QFT-action-principle}, the classical and rigorous procedures for
calculating derivatives relative to operator\ndash valued arguments coincide
if variations of the arguments proportional to the identity operator are
considered. At any rate, since in this paper particular operator derivatives
will not be computed (with a single exception), we shall treat the operator
derivatives as accepted in the literature. If a rigorous treatment is
required, the considerations and results in this paper can be ``mended''
according to the recipe given in~\cite{bp-QFT-action-principle}.



\section
[Heisenberg picture and description of interacting fields]
{Heisenberg picture and description of interacting fields}
\label{Sect2}

	In the present section, we review the general settings of canonical
quantum field theory in Heisenberg picture of motion; for details, see,
e.g.,~\cite{Bjorken&Drell,Roman-QFT,Bogolyubov&Shirkov}. In view of the
considerations in the next sections, all quantities in Heisenberg picture
will be labeled by a tilde (wave) over their (kernel) symbols.

	Let there be given a system of $n\in\field[N]$ quantum fields. The
$i^{\mathrm{th}}$ quantum field is described via a linear operator
$\tope{\varphi}_i(x)$, $x\in\base$, acting on the Hilbert space $\Hil$ of
states of the system, $\tope{\varphi}_i(x)\colon\Hil\to\Hil$. The field
operators $\tope{\varphi}_i(x)$ are supposed to be solutions of the
Euler\ndash Lagrange equation(s)
	\begin{equation}	\label{4.1}
\frac{\pd\tope{L}}{\pd\tope{\varphi}_i(x)}
-
\frac{\pd}{\pd x^\mu}
	\Bigl( \frac{\pd\tope{L}}{\pd (\pd_\mu\tope{\varphi}_i(x)) } \Bigr)
= 0 ,
	\end{equation}
where $\tope{L}$ is the Lagrangian (density) of the system.%
\footnote{~%
As accepted in the literature on quantum field
theory~\cite{Roman-QFT,Bjorken&Drell-2}, the derivatives of operator-valued
functions of operator arguments, as the ones appearing in~\eref{4.1}, are
calculated as in a case of ordinary (classical) functions of commuting
arguments with the only exception that the order of all operators should be
preserved. In most of the cases this procedure is harmless and works well,
but leads to a certain non\ndash uniqueness in the definitions of some
(conserved) quantities. For details, see~\cite{bp-QFT-action-principle}.%
}
We suppose
$\tope{L}$ to be a function, polynomial or convergent power series, of the
field operators $\tope{\varphi}_i(x)$ and their first partial derivatives
$\pd_\mu\tope{\varphi}_i(x)$.%
\footnote{~%
Most of our results admit a straightforward generalization for Lagrangians
depending on higher derivatives of the field
operators~\cite{Jaen&et_al.,Chervyakov&Nesterenko,Simon-1990}. However, we
drop the investigation of this case by the following three reasons:
	(i)~Such Lagrangians play some role as model ones, have a number of
intrinsic problems, and there are only some indications that real processes
may be described by them;
	(ii)~Such a generalization does not change anything in the ideas and
methods in our investigation and only leads to complications in the
calculations;
	(iii)~After one has on his/her disposal the results of the present
work, the mentioned generalization is only a matter of some technical details
and corresponding calculations.%
}
The Lagrangian $\tope{L}$ is also suppose to be explicitly independent of a
spacetime point $x$ at which  it is evaluated.%
\footnote{~%
This is a serious restriction. It means that only closed or translation
invariant systems are considered.%
}
So, we have
	\begin{equation}	\label{2.1}
\tope{L}
=
\tope{L} (x)
=
\tope{L} (\tope{\varphi}_i(x) , \pd_\mu\tope{\varphi}_j(x) ) .
	\end{equation}

	In the Heisenberg picture, by definition, the state vectors from the
system's Hilbert space $\Hil$ are supposed to be constant in spacetime, \ie a
vector $\ope{X}\in\Hil$ can represent a (physically realizable) state if
$\pd_\mu\ope{X}\equiv0$. The state vectors are supposed to be eigenvectors of
all commuting observables characterizing a given system of quantum fields.

	Besides the Euler-Lagrange equations~\eref{4.1}, the field operators
$\varphi_i(x)$ are supposed to satisfy the Heisenberg equations/relations of
motion, which, in the Heisenberg picture, are
	\begin{equation}	\label{6.8}
[\tope{\varphi}_i(x), \tope{P}_\mu ]_{\_}
=
\ih\frac{\pd}{\pd x^\mu} \tope{\varphi}_i(x) ,
	\end{equation}
where $\tope{P}_\mu$ are the components of the momentum operator, defined
below by~\eref{4.6}. These equations express the transformation properties of
the field operators or that the momentum operator can be considered, in a
sense, as a generator of the translation operator in the space of operators
on systems' Hilbert space of states~\cite{Bogolyubov&Shirkov}.%
\footnote{~%
For some details, see~\cite{bp-QFT-momentum-operator}.%
}
At present, there are considered only Lagrangians for which the
equations~\eref{4.1}, \eref{6.8} and~\eref{4.6} are
compatible~\cite[\S~68]{Bjorken&Drell-2}.

	The general framework of canonical quantization
describes equally well free and interacting fields. It is generally accepted,
the Lagrangian $\tope{L}$ of a system of quantum fields $\tope{\varphi}_i(x)$
to be decomposes as a sum
	\begin{equation}	\label{12.1}
\tope{L} = \tope[0\mspace{-3mu}]{L} + \tope[\prime\mspace{-3mu}]{L}
	\end{equation}
of a Lagrangian $\tope[0\mspace{-3mu}]{L}$, called \emph{free Lagrangian},
describing a system consisting of
the same fields considered as free ones and a term
$\tope[\prime\mspace{-3mu}]{L}$, called \emph{interaction Lagrangian},
describing the interaction between these fields.
Examples of free or interacting Lagrangians can be found in any (text)book on
quantum field theory, \eg
in~\cite{Bogolyubov&Shirkov,Roman-QFT,Bjorken&Drell}.%
\footnote{~%
One should always take into account the
normal ordering in $\tope{L}$ which is, normally, not indicated. For some
peculiarities of frequency decompositions and normal products of non\ndash
free fields, see, e.g.,~\cite{Bogolyubov&Shirkov,Roman-QFT}.%
}


	The decomposition~\eref{12.1} entails similar ones for the
energy-momentum tensorial operator%
\footnote{~%
The order of the operators in the first term in~\eref{4.6new1} is essential.
More generally, the first term in~\eref{4.6new1} may be defined in different
ways, which leads to different `intermediate' theories that agree after the
establishment of (anti)commutation relations and normal ordering of products
(compositions) of creation and annihilation operators. However, in this work
only some properties of the momentum operator~\eref{4.6} which are
independent of a particular definition of energy-momentum operator will be
used.

	As demonstrated in~\cite{bp-QFT-action-principle}, the rigorous
definition of a derivative with respect to operator\ndash valued argument
entails unique expressions for the conserved operators obtained via the first
Noether theorem, in particular for the energy\ndash momentum operator.%
}
	\begin{gather}	\label{4.6new1}
\tope{T}^{\mu\nu} (x)
:=
\tope{\pi}^{i\mu}(x) \circ \pd^\nu\tope{\varphi}_i(x) - \eta^{\mu\nu}\tope{L},
\intertext{where}	\label{4.6new2}
\tope{\pi}^{i\mu}(x) := \frac{\pd\tope{L}}{\pd(\pd_\mu\tope{\varphi}_i(x))} ,
	\end{gather}
and for the (canonical) 4\ndash momentum operator
	\begin{equation}	\label{4.6}
\tope{P}^\mu
:=
\frac{1}{c} \int_{\sigma} \tope{T}^{\nu\mu}(x) \Id\sigma_\nu(x) ,
	\end{equation}
where $c$ is the velocity of light in vacuum and the integration is over some
space\ndash like surface $\sigma$.
Indeed, since~\eref{4.6new2} implies
	\begin{equation}	\label{12.2}
\tope{\pi}^{i\mu}
:=
\frac{\pd\tope{L}}{\pd(\pd_\mu\tope{\varphi}_i(x))}
=
\tope[0]{\pi}^{i\mu} + \tope[\prime]{\pi}^{i\mu}
\qquad
\tope[a]{\pi}^{i\mu} = \frac{\pd\tope[a\mspace{-3mu}]{L}}
				{\pd(\pd_\mu\tope{\varphi}_i(x))}
\quad
a=0,\prime ,
	\end{equation}
from~\eref{4.6new1}, we get
	\begin{equation}	\label{12.3}
	\begin{split}
\tope{T}^{\mu\nu}(x)
&=
\tope[0]{T}^{\mu\nu}(x) + \tope[\prime]{T}^{\mu\nu}(x)
\\
\tope[a]{T}^{\mu\nu}(x)
&=
\tope[a]{\pi}^{i\mu}(x) \circ \pd^\nu\tope{\varphi}_i(x)
		- \eta^{\mu\nu} \tope[a\mspace{-3mu}]{L}(x)
\quad
a=0,\prime
	\end{split}
	\end{equation}
which, due to~\eref{4.6}, yields
	\begin{equation}	\label{12.4}
\tope{P}_{\mu}
=
\tope[0]{P}_{\mu} + \tope[\prime]{P}_{\mu}
\qquad
\tope[a]{P}_{\mu}
=
\frac{1}{c} \int \tope[a]{T}\Sprindex{\mu}{\nu}(x) \Id\sigma_\nu(x)
\quad
a=0,\prime .
	\end{equation}

	We should mention an important special case, named
\emph{nonderivative coupling}, when the interaction Lagrangian
$\tope[\prime\mspace{-3mu}]{L}$ does not depend on the derivatives
$\pd_\mu\tope{\varphi}_i(x)$ of the field operators $\tope{\varphi}_i(x)$.
In it
	\begin{equation}	\label{12.5}
\tope[\prime]{\pi}^{i\mu} = 0
\qquad
\tope{\pi}^{i\mu}
= \tope[0]{\pi}^{i\mu}
= \frac{\pd\tope[0\mspace{-3mu}]{L}} {\pd(\pd_\mu\tope{\varphi}_i(x))} ,
	\end{equation}
so that
	\begin{equation}	\label{12.6}
\tope{T}^{\mu\nu}
=
\tope[0]{T}^{\mu\nu} - \eta^{\mu\nu} \tope[\prime\mspace{-3mu}]{L} .
	\end{equation}
Hereof, by~\eref{12.4},
	\begin{equation}	\label{12.7}
\tope{P}_0
=
\tope[0]{P}_0 + \tope[\prime]{P}_0
\qquad
\tope{P}_a  = \tope[0]{P}_a	\quad   a=1,2,3 .
	\end{equation}

	The decomposition~\eref{12.3} implies a similar one for the system's
Hamiltonian $\tope{H}=\tope{T}^{00}$, \viz
	\begin{equation}	\label{12.7new}
\tope{H} = \tope[0]{H} + \tope[\prime]{H}
	\end{equation}
in which $\tope[\prime]{H}$ is called the \emph{interaction Hamiltonian}. In
a case of nonderivative coupling, in view of~\eref{12.6} and~\eref{12.7}, it
coincides up to a sign with the interaction Lagrangian, \ie
	\begin{equation}	\label{12.8}
\tope[\prime]{H} = -  \tope[\prime\mspace{-3mu}]{L} .
	\end{equation}

	On the above decompositions are based many model theories,
investigated in the literature, and are elaborated a number of specific
methods, such as ones involving in-, out-, and bare states or the
perturbation and renormalization theories.


	It is well known, the commutation relations between the field
operators and/or some functions of them and their partial derivatives play a
very important role in quantum field theory. However, rigorously speaking,
they are known only for free fields or, more precisely, when the field
operators satisfy the free Euler\ndash Lagrange equations.%
\footnote{~%
For instance, the last case is realized in the interaction picture for
nonderivative coupling between the fields; see sections~\ref{Sect3}
and~\ref{Sect4}.%
}
Since we intend to consider the problem of commutation relations in different
pictures of motion in a separate paper, it will not be dealt with in the
present work.


\section {Interaction picture. I.\ Covariant formulation}
\label{Sect3}

	The shift from Heisenberg picture, summarized in Sect.~\ref{Sect2},
to the interaction one is realize by means of the so\ndash called
$\ope{U}$\ndash operator
which is connected with the interaction Hamiltonian in a way similar to the
one the evolution operator in non-relativistic quantum mechanics is linked to
the Hamiltonian~\cite{Prugovecki-QMinHS,bp-BQM-introduction+transport}. In
this sense, the $\ope{U}$\ndash operator plays a role of evolution operator
in the (interaction picture of) quantum field theory. But such a view\ndash
point is limited as the absence of interaction, which entails
$\ope{U}=\id_\Hil$, does not mean a complete disappearance of (\eg time)
evolution of the free fields.%
\footnote{~%
However, since the main aim of quantum field theory is the description of
interacting (quantum) fields and/or elementary particles, we may write the
pure symbolical equality \emph{evolution=interaction} and treat  $\ope{U}$ as
evolution operator. (The fibre bundle treatment of this problem will give
other arguments in favor of such understanding; it will be developed
elsewhere.)%
}

	We start with the so-called covariant approach which utilizes the
notion of a functional derivative with respect to a (space-like in our
context) surface $\sigma$ in $\base$,%
\footnote{~%
The only advantage of that approach is its explicit covariance. However, in
our opinion, it is, in some sense, complicated and does not bring much to the
understanding of the geometrical structure of quantum field theory. Besides,
it essentially uses the specific properties of Minkowski spacetime, thus
making the generalization of quantum field theory on curved manifolds more
difficult.%
}

	Recall, if $G\colon\sigma\mapsto G[\sigma]\in V$, $V$ being a vector
space (\eg $V=\field[C]$ in our context below), is a functional of a
3\ndash dimensional surface $\sigma$ in the (4\ndash dimensional) Minkowski
space $\base$, the \emph{functional derivative of $G$ with respect to $\sigma$
at a point $x\in\sigma$} is a mapping
$(G,\sigma,x)\to\frac{\delta G[\sigma]}{\delta\sigma(x)}\in V$, $x\in\sigma$,
such that
	\begin{equation}	\label{12.9}
	\begin{split}
&
\frac{\delta G[\sigma]}{\delta\sigma(x)}
:=
\frac{\delta G}{\delta\sigma}\Big|_{x}
:=
\lim_{ \substack{\Omega\to\{x\} \\ \mathrm{vol}(\Omega)\to 0} }
\frac{G[\sigma'] - G[\sigma] } {\mathrm{vol}(\Omega)}
\\
&
\sigma':= (\sigma\backslash(\pd\Omega\cap\sigma))
  \cup (\pd\Omega\backslash(\pd\Omega\cap\sigma))
	\end{split}
	\end{equation}
if the limit in the r.h.s.\ exists. Here $\Omega$ is a (closed) 4-dimensional
submanifold of $\base$ with boundary, $\pd\Omega$ is its boundary, and
$\mathrm{vol}(\Omega)$ is the (4\ndash dimensional) volume of $\Omega$.
Besides, it is supposed that the intersection $\Omega\cap\sigma$ is not empty,
contains the point $x$, lies in $\pd\Omega$, and is a 3\ndash dimensional
submanifold of $\pd\Omega$ without boundary.%
\footnote{~%
All this means that the surface $\sigma'$ is obtained from $\sigma$ via a
continuous deformation in a neighborhood of $x$ with $\Omega$ being the
spacetime region between $\sigma$ and $\sigma'$.%
}

	For example, we have:~\cite[pp.~9--11]{Roman-QFT}
	\begin{gather}	\label{12.10}
\frac{\delta}{\delta\sigma(x)} 	\int_\sigma f^\mu(y) \Id\sigma_\mu(y)
=
\pd_\mu f^\mu(x)
\\			\label{12.11}
\frac{\delta}{\delta\sigma(x)} 	\int_\sigma f(y) \Id\sigma_\mu(y)
=
\pd_\mu f(x),
	\end{gather}
where $f^\mu,f\colon\base\to\field[C]$ are of class $C^1$. (The second
equality follows from the first one for $f^\nu=\delta_{\mu}^{\nu}f$.)

	The equality $\frac{\delta G[\sigma]}{\delta\sigma}=0$ is a
criterion for surface\ndash independence of $G$. In the context of
quantum field theory, we put $V=\field[C]$ and the surface $\sigma$ will
always be assumed \emph{space\ndash like}, \ie its normal vector $n^\mu$ is
supposed time\ndash like, $n_\mu n^\mu=+1$ (or, more generally
$n_\mu n^\mu>0$) everywhere on $\sigma$, or, equivalently, if
$x,y\in\sigma$ with $x\not=y$, then $(x-y)^2<0$. In this case, the functional
derivative~\eref{12.9} is a generalization of the partial time derivative.
Indeed, if $\sigma$ is constant time surface, \ie
$\sigma=\{x\in\base : x^0=ct=\const\}$, then%
\footnote{~%
The same result holds if the elementary regions are scale from
flat~\cite{Kanesawa&Tomonaga}.%
}
	\begin{equation}	\label{12.12}
\frac{\pd f(x)}{\pd x^0}
=
\frac{\delta}{\delta\sigma(x)}
	\int\limits_{\sigma_0}^{\sigma} f(y)\Id y
=
\frac{\delta}{\delta\sigma(x)}
	\int\limits_{\sigma_0}^{\sigma} f(y)|_{y\in\sigma'}\Id\sigma'
	\end{equation}
where the integrals mean that one, at first, integrates along a surface
$\sigma'$ from a foliation $\Sigma$ of surfaces `between'  $\sigma_0$ and
$\sigma$ and then along (some parameter, possibly the time, characterizing)
the foliation $\Sigma$. One should compare~\eref{12.12} with~\eref{12.11} for
a constant time surface and $\Id\sigma_\mu(y)=\delta_{\mu}^{0}\Id^3\bs y$,
that is with
	\begin{equation}	\label{12.13}
\frac{\delta}{\delta\sigma(x)}
	\int\limits_{y^0=ct=\const}^{} f(y)\Id^3\bs{y}
=
\frac{\pd f(x)}{\pd x^0} .
	\end{equation}

	The basic idea of the interaction picture is to chose
$\ope{V}$ in~\eref{12.14} in such a way that in it to be incorporated the
information about the interaction of the fields. It is realized a little
below. Considerations of the interaction picture of quantum field theory
with \emph{nonderivative} coupling can be found in any serious (text)book on
quantum field
theory~\cite{Bogolyubov&Shirkov,Roman-QFT,Bjorken&Drell}. However, the
reading or early works,
like~\cite{Tomonaga,Schwinger-E1,Schwinger-E2,Schwinger-E3,
Schwinger-QFT-1,Schwinger-QFT-2,Schwinger-QFT-3,QED-1954}, can help much.
Besides, in works
like~\cite{Schwinger-E1,Matthews-1,Matthews-2,Rohrlich,Lee&Yang} one can find
consideration of/or methods applicable to investigation of interaction
Lagrangians/Hamiltonians with \emph{derivative} coupling which is absent in
most (text)books.

	Let $\Sigma$ be a 3-dimensional foliation of the spacetime $\base$
consisting of space\ndash like surfaces, $\sigma_0\in\Sigma$ be arbitrarily
fixed, and $\sigma\in\Sigma$.%
\footnote{~%
On the foliation theory, see, e.g.,~\cite{Tamura,Hector&Hirsch-1}. It is
essential to be noted, the space\ndash like surfaces of a given foliation of
Minkowski space transform into each other under a Lorentz transformation,
forming a group~\cite{Baumann-1,Baumann-2}.%
}
Define a unitary operator $\ope{U}[\sigma,\sigma_0]\colon\Hil\to\Hil$, which
is an operator\ndash valued functional of $\sigma$ and $\sigma_0$, as the
unique solution of the Tomonaga\ndash Schwinger equation
	\begin{equation}	\label{12.16}
\ih c \frac{\delta\ope{U}[\sigma,\sigma_0]}{\delta\sigma(x)}
=
\ope[\prime]{H}[\sigma;x]\circ \ope{U}[\sigma,\sigma_0]
\quad
x\in\sigma
	\end{equation}
satisfying the initial condition
	\begin{equation}	\label{12.17}
\ope{U}[\sigma_0,\sigma_0] = \id_\Hil .
	\end{equation}
Here $\ope[\prime]{H}[\sigma;x]$ is the interaction Hamiltonian density, \ie
the interaction energy density, defined a little below by~\eref{12.18c}.
The unitarity of $\ope{U}[\sigma,\sigma_0]$ means that
$\ope{U}^\dag[\sigma,\sigma_0] = \ope{U}^{-1}[\sigma_0,\sigma]$, where
the Hermitian conjugate $\ope{U}^\dag$ of $\ope{U}$ is defined via
	\begin{equation}	\label{12.19}
\clangle \ope{Y}[\sigma] |
		\ope{U}[\sigma,\sigma_0] (\ope{X}[\sigma]) \crangle
=
\clangle \ope{U}^\dag[\sigma_0,\sigma] (\ope{Y}[\sigma]) |
					\ope{X}[\sigma] \crangle
	\end{equation}
for every $\Hil$\ndash valued functionals $\ope{X}[\sigma]$ and
$\ope{Y}[\sigma]$.%
\footnote{~%
In the literature the arguments of $\ope{U}^\dag$ are interchanged, \ie
 $\ope{U}^\dag[\sigma_0,\sigma]$ in our notation is denoted as
$\ope{U}^\dag[\sigma,\sigma_0]$, which we find inconvenient and not suitable
for our system of notation.%
}

	It is said that the transformations
	\begin{subequations}	\label{12.18}
	\begin{align}		\label{12.18a}
\tope{X}\mapsto \ope{X}[\sigma]
	&:= \ope{U}[\sigma,\sigma_0]  (\tope{X})
\\				\label{12.18b}
\tope{A}(x)\mapsto \ope{A}[\sigma;x]
	&:= \ope{U}[\sigma,\sigma_0] \circ \tope{A}(x)\circ
	    \ope{U}^{-1}[\sigma,\sigma_0]
\qquad
x\in\sigma
\intertext{realize the \emph{transition to the interaction picture} if
 $\ope{U}[\sigma,\sigma_0]$ is defined via the equation~\eref{12.16} under
the initial condition~\eref{12.17} with}
				\label{12.18c}
\ope[\prime]{H}[\sigma;x]
	&:= \ope{U}[\sigma,\sigma_0] \circ \tope[\prime]{H}(x) \circ
   	    \ope{U}^{-1}[\sigma,\sigma_0]
\quad
x\in\sigma
	\end{align}
	\end{subequations}
where $\tope[\prime]{H}(x)$ is the interaction Hamiltonian in Heisenberg
picture.
	The description of a quantum system via the ($\Hil$\ndash valued)
functionals $\ope{X}$, representing the system's states and called state
functionals, and (operator\ndash valued and, possibly, point dependent)
functionals $\ope{A}[\sigma;x]$, called operator functionals, is called the
\emph{interaction picture} or \emph{interaction representation} of the
(motion) of the considered system. In particular, in this picture the field
operators $\tope{\varphi}_i(x)$ and an operator $\tope{A}[\sigma_0]$ given on
$\sigma_0$ are represented by the functionals
	\begin{align}		\label{12.20}
\varphi_i[\sigma;x]
	&:= \ope{U}[\sigma,\sigma_0] \circ \tope{\varphi}_i(x)\circ
	    \ope{U}^{-1}[\sigma,\sigma_0]
\\				\label{12.21}
\ope{A}[\sigma]
	&:= \ope{U}[\sigma,\sigma_0] \circ \tope{A}[\sigma_0]\circ
	    \ope{U}^{-1}[\sigma,\sigma_0] .
	\end{align}

	Before going ahead, we want to make some comments on the above
definitions which, we hope, will contribute to their better understanding.

	In~\eref{12.18c} enters the interaction Hamiltonian
$\tope[\prime]{H}(x)$ in Heisenberg picture. In accord with~\eref{12.4}
it is (cf.~\cite[eq.~(4)]{Matthews-1})
	\begin{equation}	\label{12.22}
\tope[\prime]{H}(x) = \tope[\prime]{T}^{\mu\nu}(x) n_\mu(x) n_\nu(x)
	\end{equation}
where $n_\mu(x)$ is the unit normal vector to $\sigma$ at $x\in\sigma$. The
particular choice of $\sigma$ as constant time surface implies
$n_\mu(x)=\delta_\mu^0$, so that
	\begin{equation}	\label{12.23}
\tope[\prime]{H}(x)
= \tope[\prime]{T}^{00}(x)
= \tope[0]{\pi}^{i0} \circ \pd^0\! \tope{\varphi}_i(x) - \tope{L}(x)
	\end{equation}
which in a case of nonderivative coupling reduces to~\eref{12.8}, due
to~\eref{12.5}.

	Obviously (see~\eref{12.18b}), the functional $\ope[\prime]{H}$
in~\eref{12.16} is the Hamiltonian density in the interaction picture. It is
supposed to be know when one works in the interaction picture. However, if
one knows it in the Heisenberg picture, then the actual equation for
$\ope{U}$ is obtained from~\eref{12.16} by inserting in it~\eref{12.18c}
which, in view of
	\begin{align}	\label{12.26}
\ope{U}^\dag[\sigma_0,\sigma]
& =
\ope{U}^{-1}[\sigma,\sigma_0] ,
\\ \intertext{expressing the unitarity of $\ope{U}$, results in}
			\label{12.24}
\ih c \frac{\delta\ope{U}[\sigma,\sigma_0]}{\delta\sigma(x)}
& =
\ope{U}[\sigma,\sigma_0] \circ \tope[\prime]{H}(x)
\quad
x\in\sigma
\\ \intertext{or, equivalently,}
			\label{12.25}
-\ih c \frac{\delta\ope{U}^{-1}[\sigma,\sigma_0]}{\delta\sigma(x)}
& =
\tope[\prime]{H}(x)\circ \ope{U}^{-1}[\sigma,\sigma_0]
\quad
x\in\sigma .
	\end{align}

	Since the solutions of~\eref{12.16} (or~\eref{12.24},
or~\eref{12.25}) satisfy the equalities
	\begin{align}	\label{12.27}
\ope{U}^{-1}[\sigma,\sigma_0] &= \ope{U}[\sigma_0,\sigma]
\\			\label{12.28}
\ope{U}[\sigma,\sigma_0]
	&= \ope{U}[\sigma,\sigma_1]\circ \ope{U}[\sigma_1,\sigma_0] ,
	\end{align}
we can rewrite~\eref{12.25} also as
	\begin{equation}	\label{12.29}
-\ih c \frac{\delta\ope{U}[\sigma_0,\sigma]}{\delta\sigma(x)}
=
\tope[\prime]{H}(x)\circ \ope{U}[\sigma_0,\sigma]
\quad
x\in\sigma .
	\end{equation}
It is clear, the equations~\eref{12.24}, \eref{12.25}, and~\eref{12.29} are
equivalent to the basic Tomonaga\ndash Schwinger equation~\eref{12.16}.

	If the interaction Hamiltonians $\ope[\prime]{H}[\sigma;x]$ commute
on arbitrary surfaces, \ie
	\begin{equation}	\label{12.29new}
[ \ope[\prime]{H}[\sigma;x] , \ope[\prime]{H}[\sigma';x'] ]_{\_} = 0
\qquad	x\in\sigma
\quad	x'\in\sigma' ,
	\end{equation}
the last conclusion is also a consequence of the equality
	\begin{equation}	\label{12.29new1}
\ope[\prime]{H}[\sigma;x] = \tope[\prime]{H}(x)
	\end{equation}
which, in turn, is a corollary of~\eref{12.18c} and the commutativity of
$\ope{U}[\sigma,\sigma_0]$ and $\ope[\prime]{H}[\sigma;x]$,
	\begin{equation}	\label{12.29new2}
[ \ope{U}[\sigma,\sigma_0] , \ope[\prime]{H}[\sigma;x] ]_{\_} = 0.
	\end{equation}
(The last fact is a consequence of equation~\eref{12.16} (see also the
`explicit' solution~\eref{12.54} of~\eref{12.16} presented below).)

	One may ask about the integrability conditions of~\eref{12.16} with
respect to $\ope{U}$ considered as two\ndash argument functional, \ie is the
equality
\(
\frac{\delta^2\ope{U}[\sigma,\sigma_0]}{\delta\sigma(x)\delta\sigma_0(y)}
=
\frac{\delta^2\ope{U}[\sigma,\sigma_0]}{\delta\sigma_0(y)\delta\sigma(x)}
\)
for $x\in\sigma$ and $y\in\sigma_0$ valid or not? The easiest way to check
this is the second functional derivatives to be computed by means
of~\eref{12.24} and~\eref{12.29}. The result is
\[
\frac{\delta^2\ope{U}[\sigma,\sigma_0]}{\delta\sigma(x)\delta\sigma_0(y)}
=
\frac{\delta^2\ope{U}[\sigma,\sigma_0]}{\delta\sigma_0(y)\delta\sigma(x)}
=
- \frac{1}{(\ih c)^2}
\tope[\prime]{H}(y) \circ\ope{U}[\sigma,\sigma_0]\circ \tope[\prime]{H}(x) .
\]
Consequently~\eref{12.16} is always integrable. Here we want to point to an
error in~\cite[p.~161]{Roman-QFT} where it is stated that the considered
integrability condition should be
	\begin{equation}	\label{12.30}
[\tope[\prime]{H}(x) , \tope[\prime]{H}(y)]_{\_} = 0
\qquad
x,y\in\sigma .
	\end{equation}
Since~\eref{12.24} implies
	\begin{equation}	\label{12.31}
\frac{\delta^2\ope{U}[\sigma,\sigma_0]}{\delta\sigma(x)\delta\sigma(y)}
=
\frac{1}{(\ih c)^2}
\ope{U}[\sigma,\sigma_0]\circ \tope[\prime]{H}(x)\circ \tope[\prime]{H}(y)
\qquad
x,y\in\sigma ,
	\end{equation}
we see that~\eref{12.30} is tantamount to
	\begin{equation}	\label{12.32}
\frac{\delta^2\ope{U}[\sigma,\sigma_0]}{\delta\sigma(x)\delta\sigma(y)}
=
\frac{\delta^2\ope{U}[\sigma,\sigma_0]}{\delta\sigma(y)\delta\sigma(x)}
\qquad
x,y\in\sigma
	\end{equation}
which is an \emph{additional} condition on $\ope{U}$, that may or may not
hold, not an integrability conditions of~\eref{12.16}. We should note, the
pointed error in~\cite[p.~161]{Roman-QFT} is harmless as in this
book~\eref{12.30} is identically valid under the conditions assumed in
\emph{loc.\ cit.}

	It should be stressed, in the most cases the interaction operators
(functionals), as defined by~\eref{12.18b} (and~\eref{12.18c} for the
Hamiltonian), are independent of the spacelike surface $\sigma\in\Sigma$ and
depend only on the point $x$ in a sense that
	\begin{equation}	\label{12.33}
\frac{\delta\ope{A}[\sigma;x]}{\delta\sigma(y)} = 0
\qquad
\text{ for $(x-y)^2<0$ (\ie  $x,y\in\sigma$, $x\not=y$)}
	\end{equation}
if $\ope{A}[\sigma;x]$ and $\ope{H}[\sigma;x]$ (or $\tope{A}(x)$ and
$\tope[\prime]{H}(x))$ are polynomial or convergent power series in the field
functionals~\eref{12.20} (or field operators $\varphi_i(x)$) in which the
fermion fields, if any, enter only in terms of even degree relative to them
(counting every fermion field component in it).%
\footnote{~%
Proof: From~\eref{12.18b} and~\eref{12.16}, we get
\(
\frac{\delta\ope{A}[\sigma;x]}{\delta\sigma(y)}
=
\frac{\delta\ope{U}[\sigma,\sigma_0]}{\delta\sigma(y)}
	\circ\tope{A}(x)\circ \ope{U}^\dag[\sigma_0,\sigma]
+
	\ope{U}[\sigma,\sigma_0]\circ\tope{A}(x)\circ
\frac{\delta\ope{U}^{-1}[\sigma,\sigma_0]}{\delta\sigma(y)}
=
\frac{1}{\ih c} [\ope{H}[\sigma;x],\ope{A}[\sigma;x] ]_{\_}
=
\ope{U}[\sigma,\sigma_0]\circ
\frac{1}{\ih c} [\tope{H}(x),\tope{A}(x) ]_{\_}
\circ\ope{U}^\dag[\sigma_0,\sigma]
= 0
\)
as, under the conditions specified, $[\tope{H}(x),\tope{A}(x) ]_{\_}=0$ in
view of the (anti)commutation relations
 $[\tope{\varphi}_i(y), \tope{\varphi}_i(x)]_\pm=0$ for $(x-y)^2<0$ (the
sign plus (minus) corresponds to fermion (boson) fields); the last result is a
consequence of the linearity of the (anti)commutator and the
identities~\eref{12.48} given below.%
}
In particular, for a polynomial Hamiltonian with nonderivative coupling, we
have
	\begin{equation}	\label{12.35}
\frac{\delta\varphi_i[\sigma;x]}{\delta\sigma(y)} = 0
\qquad
(x-y)^2 < 0  \quad x,y\in\sigma .
	\end{equation}

	We turn now our attention to the equations of motion in the
interaction picture. In contrast to most (text)books, such
as~\cite{Bogolyubov&Shirkov,Bjorken&Drell,Roman-QFT}, we consider arbitrary
interactions, with or without derivative coupling.

	The Tomonaga-Schwinger equation~\eref{12.16} replaces the
Schr\"odinger one of non-relativistic quantum mechanics in the
interaction picture of quantum field theory and plays a role of a dynamical
equation of motion for the states (state functionals). Indeed,
combining~\eref{12.16} and~\eref{12.18a}, we get
	\begin{align}	\label{12.36}
\ih c \frac{\delta\ope{X}[\sigma]}{\delta\sigma(x)}
=
\ope[\prime]{H}[\sigma;x] (\ope{X}[\sigma])
\qquad
x\in\sigma
\intertext{with initial condition (see~\eref{12.17})}
			\label{12.37}
\ope{X}[\sigma]|_{\sigma=\sigma_0} = \ope{X}[\sigma_0] = \tope{X} .
	\end{align}
We emphasize, a state functional $\ope{X}[\sigma]$ depends on the spacelike
surface $\sigma$ as a whole, not on a particular point(s) in it, which is not
the case when observables and other functionals, like $\ope{A}[\sigma;x]$,
are explored as they essentially depend on $x\in\sigma$ and, under afore
given conditions, are independent of $\sigma$ in a sense of~\eref{12.33}.

	To derive the equations of motion for field functionals, we shall
generalize the method used in~\cite[sec.~2, eq.~(2.7)--(2.11)]{Schwinger-E1}
for the same purpose but in quantum electrodynamics with ordinary non\ndash
derivative coupling.%
\footnote{~%
We present the derivation of the equations of motion since the author of
these lines fails to find these equations in the general case, possibly
containing derivative coupling, in the available to him literature;
implicitly (and in special coordinates) they are contained
in~\cite[eq.~(5)]{Matthews-1}, but these are not the \emph{explicit} equations
we need.%
}
From~\eref{12.11} with $\ope{A}[\sigma;x]$ for $f(x)$, we find
	\begin{multline*}
\frac{\pd\ope{A}[\sigma;x]}{\pd x^\mu}
=
\frac{\delta}{\delta\sigma(x)} \int_\sigma \ope{A}[\sigma;y] \Id\sigma_\mu(y)
\\
=
\frac{\delta}{\delta\sigma(x)}
	\Bigl\{ \ope{U}[\sigma,\sigma_0]
	\circ \int_\sigma \tope{A}(y) \Id\sigma_\mu(y) \circ
	\ope{U}^\dag[\sigma_0,\sigma] \Bigr\}
\\
=
\frac{\delta\ope{U}[\sigma,\sigma_0]}{\delta\sigma(x)}
	\circ \int_\sigma \tope{A}(y) \Id\sigma_\mu(y) \circ
	\ope{U}^\dag[\sigma_0,\sigma]
+
\ope{U}[\sigma,\sigma_0] \circ
	\Bigl\{ \frac{\delta}{\delta\sigma(x)}
	\int_\sigma \tope{A}(y) \Id\sigma_\mu(y)  \Bigr\}
\\\circ
	\ope{U}^\dag[\sigma_0,\sigma]
+
\ope{U}[\sigma,\sigma_0]
	\circ \int_\sigma \tope{A}(y) \Id\sigma_\mu(y) \circ
	\frac{\delta}{\delta\sigma(x)}\ope{U}^{-1}[\sigma,\sigma_0] .
	\end{multline*}
Expressing
$\frac{\delta\ope{U}^{\pm1}[\sigma,\sigma_0]}{\delta\sigma(x)}$
from~\eref{12.16} and applying, again,~\eref{12.11}, we deduce the equality
	\begin{equation}	\label{12.38}
\ih c \frac{\pd\ope{A}[\sigma;x]}{\pd x^\mu}
=
\ope{U}[\sigma,\sigma_0]
\circ \ih c\frac{\pd\tope{A}(x)}{\pd x^\mu} \circ
\ope{U}^\dag[\sigma_0,\sigma]
+ \Bigl[\ope[\prime]{H}[\sigma;x] ,
	\int_\sigma \ope{A}[\sigma;y] \Id\sigma_\mu(y) \Bigr]_{\_} .
	\end{equation}

	Let
$\tope{L}=\tope{L}\bigl(\tope{\varphi}_i(x),\pd_\mu\tope{\varphi}_j(x)\bigr)$
be the Lagrangian of a system of quantum fields $\tope{\varphi}_i(x)$.
Suppose $\tope{L}$ is a polynomial or convergent power series in its
arguments $\tope{\varphi}_i(x)$ and  $\pd_\mu\tope{\varphi}_j(x)$. In the
interaction picture the Lagrangian functional is
	\begin{multline}	\label{12.39}
\ope{L}\bigl[
	\sigma,\sigma_0;
	\ope{\varphi}_i[\sigma;x],\pd_\mu\ope{\varphi}_j[\sigma;x]
	\bigr]
=
\ope{U}[\sigma,\sigma_0] \circ
\tope{L}\bigl(\tope{\varphi}_i(x),\pd_\mu\tope{\varphi}_j(x)\bigr)
\circ  \ope{U}^\dag[\sigma_0,\sigma]
\\
=
\tope{L}\bigl(
	\ope{\varphi}_i[\sigma;x],
	\ope{U}[\sigma,\sigma_0]
		\circ \pd_\mu\tope{\varphi}_j(x) \circ
	\ope{U}^\dag[\sigma_0,\sigma]
	\bigr) .
	\end{multline}
Since~\eref{12.38} with $\tope{A}(x)=\tope{\varphi}_j(x)$ yields
	\begin{equation}	\label{12.40}
\frac{\pd\varphi_j[\sigma;x]}{\pd x^\mu}
=
\ope{U}[\sigma,\sigma_0]
\circ \ih c\frac{\pd\tope{\varphi}_i(x)}{\pd x^\mu} \circ
\ope{U}^\dag[\sigma_0,\sigma]
+ \frac{1}{\ih c} \Bigl[\ope[\prime]{H}[\sigma;x] ,
	\int_\sigma \varphi_j[\sigma;y] \Id\sigma_\mu(y) \Bigr]_{\_} ,
	\end{equation}
we see that the explicit functional form of $\ope{L}$ is
	\begin{equation}	\label{12.41}
\ope{L}
=
\tope{L}\bigl(
	\ope{\varphi}_i[\sigma;x],
	\pd_\mu\ope{\varphi}_i[\sigma;x]
	- \frac{1}{\ih c} \Bigl[\ope[\prime]{H}[\sigma;x] ,
	\int_\sigma \varphi_j[\sigma;y] \Id\sigma_\mu(y) \Bigr]_{\_}
	 \bigr) .
	\end{equation}

	Now the idea is for $\varphi_i[\sigma;x]$ to be obtained an
Euler-Lagrange type equations with, generally, non-vanishing r.h.s. For the
purpose, we intend to apply~\eref{12.38} to
 $\tope{A}(x)=\frac{\pd\tope{L}}{\pd( \pd_\mu\tope{\varphi}_i(x) )}$
and then to sum over $\mu$. On one hand, in view of the Euler\ndash Lagrange
equation~\eref{4.1}, the first term in the r.h.s.\ of~\eref{12.38} will,
after performing the described procedure, give%
\footnote{~%
Since $\ope{U}[\sigma,\sigma_0]$ does not depend on
$\tope{\varphi}_i(x)$ and $\varphi_i[\sigma;x]$ and their partial derivatives
as separate arguments, we freely move $\ope{U}[\sigma,\sigma_0]$ and
$\ope{U}^\dag[\sigma_0,\sigma]=\ope{U}^{-1}[\sigma,\sigma_0]$ through the
partial derivative operators
 $\frac{\pd}{\pd\tope{\varphi}_i(x)}$,
 $\frac{\pd}{\pd\varphi_i[\sigma;x]}$,
 $\frac{\pd}{\pd(\pd_\mu\tope{\varphi}_i(x))}$, and
 $\frac{\pd}{\pd(\pd_\mu\varphi_i[\sigma;x])}$.%
}
\[
\ih c
\ope{U}[\sigma,\sigma_0]
	\circ \frac{\pd\tope{L}}{\pd\tope{\varphi}_i(x)} \circ
\ope{U}^\dag[\sigma_0,\sigma]
=
\ih c \frac{\pd\ope{L}}{\pd\tope{\varphi}_i(x)}
=
\ih c \frac{\pd\ope{L}}{\pd\ope{\varphi}_i[\sigma;x]} .
\]
On other hand, using~\eref{12.39} and~\eref{12.40}, we derive:
	\begin{multline*}
\ope{A}[\sigma;x]
 =
\ope{U}[\sigma,\sigma_0]
	\circ \frac{\pd\tope{L}}{\pd( \pd_\mu\tope{\varphi}_i(x) )} \circ
\ope{U}^\dag[\sigma_0,\sigma]
 =
\frac{\pd\ope{L}}{\pd( \pd_\mu\tope{\varphi}_i(x) )}
 =
\frac{\pd\ope{L}}{\pd( \pd_\nu\ope{\varphi}_j[\sigma;x] )}
\\
\circ
\frac{\pd( \pd_\nu\ope{\varphi}_j[\sigma;x] ) }
     {\pd( \pd_\mu\tope{\varphi}_i(x) )}
 =
\frac{\pd\ope{L}}{\pd( \pd_\nu\ope{\varphi}_j[\sigma;x] )}
	\circ
\Bigl\{
	\ope{U}[\sigma,\sigma_0]
	\circ(\delta_\nu^\mu\delta_j^i\id_\Hil) \circ
	\ope{U}^\dag[\sigma_0,\sigma]
\\
+
\frac{1}{\ih c}
	\Bigl[
    \frac{\pd\ope[\prime]{H}[\sigma;x]}  {\pd( \pd_\mu\tope{\varphi}_i(x) )}
	,
    \int_\sigma \varphi_j[\sigma;y] \Id\sigma_\nu(y)
	\Bigr]_{\_}
\Bigr\}
 =
\frac{\pd\ope{L}}{\pd( \pd_\mu\ope{\varphi}_i[\sigma;x] )}
+ \ope{G}^{i\mu}[\sigma;x] ,
	\end{multline*}
where we have set%
\footnote{~%
An attempt to calculate
\(
\frac{\pd\ope[\prime]{H}[\sigma;x]}{\pd( \pd_\mu\tope{\varphi}_i(x) )}
=
\frac{\pd\ope[\prime]{H}[\sigma;x]}{\pd( \pd_\nu\ope{\varphi}_j[\sigma;x] )}
\frac{\pd( \pd_\nu\ope{\varphi}_j[\sigma;x] )}
     {\pd( \pd_\mu\tope{\varphi}_i(x) )}
\)
by using~\eref{12.40} results only in an iterative procedure. In actual
calculation, one should write this expression as
\(
\ope{U}[\sigma,\sigma_0] \circ
\frac{\pd\tope[\prime]{H}[\sigma;x]}{\pd( \pd_\mu\tope{\varphi}_i(x) )}
\circ \ope{U}^\dag[\sigma_0,\sigma],
\)
calculate the derivative in this formula, and, after this is done, all
quantities should be expressed in the interaction picture by means
of~\eref{12.18} and~\eref{12.40}.%
}
	\begin{multline}	\label{12.42}
\ope{G}^{i\mu}[\sigma;x]
=
\ope{G}^{i\mu}\bigl( \varphi_i[\sigma;x],\pd_\mu\varphi_j[\sigma;x] \bigr)
\\
:=
\frac{1}{\ih c}
\frac{\pd\ope{L}}{\pd( \pd_\nu\ope{\varphi}_j[\sigma;x] )}
	\circ
	\Bigl[
    \frac{\pd\ope[\prime]{H}[\sigma;x]}  {\pd( \pd_\mu\tope{\varphi}_i(x) )}
	,
    \int_\sigma \varphi_j[\sigma;y] \Id\sigma_\nu(y)
	\Bigr]_{\_}  .
	\end{multline}
Substituting the above results into afore described procedure
with~\eref{12.38}, we, finally, get:
	\begin{multline}	\label{12.43}
\frac{\pd\ope{L}}{\pd\varphi_i[\sigma;x]}
-
\frac{\pd}{\pd x^\mu}
\frac{\pd\ope{L}}{\pd( \pd_\mu\varphi_i[\sigma;x] )}
 =
\frac{\pd\ope{G}^{i\mu}[\sigma;x]}{\pd x^\mu}
\\ -
\frac{1}{\ih c}
\Bigl[
    \ope[\prime]{H}[\sigma;x]
    ,
\int_\sigma \Bigl(
	\frac{\pd\ope{L}(y)} {\pd( \pd_\mu\ope{\varphi}_i[\sigma;y] )}
	+ \ope{G}^{i\mu}[\sigma;y]
\Bigr) \Id\sigma_\mu(y)
	\Bigr]_{\_} .
	\end{multline}
These are the \emph{Euler-Lagrange equations of motion for the quantum fields
$\varphi_i[\sigma;x]$ in the (covariant) interaction picture} under the only
condition that the Lagrangians, including the `free' and `interaction' ones,
are polynomial or convergent power series in the fields and their first
partial derivatives.

	Let us now look on~\eref{12.43} in the case of \emph{nonderivative}
coupling. In it, in view of~\eref{12.2}--\eref{12.8} and~\eref{12.18},
\(
\frac{\pd\ope[\prime]{H}[\sigma;x]}{\pd( \pd_\mu\tope{\varphi}_i(x) )}
\equiv 0
\), so that
	\begin{equation}	\label{12.44}
\ope{G}^{i\mu}[\sigma;x]\big
|_{\begin{subarray}{l}\text{nonderivative}\\ \text{coupling} \end{subarray} }
\equiv 0.
	\end{equation}
Besides, due to~\eref{12.1} and~\eref{12.2}--\eref{12.8}, the equations of
motion~\eref{12.43} take the form
	\begin{equation}	\label{12.45}
\frac{\pd\ope[0\mspace{-3mu}]{L}}{\pd\varphi_i[\sigma;x]}
-
\frac{\pd}{\pd x^\mu}
\frac{\pd\ope[0\mspace{-3mu}]{L}}{\pd( \pd_\mu\varphi_i[\sigma;x] )}
 =
\frac{\pd\ope[\prime]{H}[\sigma;x]}{\pd\ope{\varphi}_i[\sigma;x] }
-
\frac{1}{\ih c}
\Bigl[
    \ope[\prime]{H}[\sigma;x]
    ,
\int_\sigma
	\frac{\pd\ope[0\mspace{-3mu}]{L}(y)}
	     {\pd( \pd_\mu\ope{\varphi}_i[\sigma;y] )}
\Id\sigma_\mu(y)
	\Bigr]_{\_} .
	\end{equation}
A few lines below it will be proved that the r.h.s.\  of this equation
identically vanishes when in $\ope[\prime]{H}$
($=-\ope[\prime\mspace{-3mu}]{L}$) enter only terms of even degree with
respect to fermion fields, if any. This means that~\eref{12.43} for
nonderivative coupling and Lagrangians of the type described reduces to
	\begin{equation}	\label{12.46}
\frac{\pd\ope[0\mspace{-3mu}]{L}}{\pd\varphi_i[\sigma;x]}
-
\frac{\pd}{\pd x^\mu}
\frac{\pd\ope[0\mspace{-3mu}]{L}}{\pd( \pd_\mu\varphi_i[\sigma;x])}
=
0 .
	\end{equation}
Thus, we have derive a well-known result: in the interaction picture the
quantum fields, \ie their functionals in our context, are solutions of the
 Euler\ndash Lagrange equations for the \emph{free} Lagrangian. The
consequences of this fundamental result are describe at length in the
literature, \eg in~\cite{Roman-QFT,Bjorken&Drell}.

	The proof of the vanishment of the r.h.s.\ of~\eref{12.45} is based
on the covariant formulation of the canonical (anti)commutation
relations,%
\footnote{~%
Let us recall them~\cite{Roman-QFT}:
	\begin{subequations}	\label{4.5}
	\begin{align}
				\label{4.5a}
[\varphi_i(x),\varphi_j(x')]_{\pm} &= 0
\qquad \text{for }(x-x')^2<0
\\				\label{4.5b}
[\pi_i(x),\pi^j(x')]_{\pm}  &= 0
\qquad \text{for }(x-x')^2<0
\\				\label{4.5c}
[\varphi_i(x),\pi^j(x')]_{\pm}   &=
		\ih\delta_{i}^{j}\delta^4(x- x')
\qquad \text{for }(x-x')^2<0
	\end{align}
	\end{subequations}
where $\pi^i:=\pi^{i0}$.%
}
viz.%
\footnote{~%
For simplicity, we suppose $\pi^{i\mu}\not=0$, \ie the system is without
constraints.%
}
	\begin{equation}	\label{12.47}
\int_\sigma
	\bigl[ \varphi_i[\sigma;x],\pi^{j\mu}[\sigma;y] \bigr]_{\pm}
\Id\sigma_\mu(y)
=
\ih c \delta_i^j ,
	\end{equation}
and on the identities
	\begin{subequations}	\label{12.48}
	\begin{align}		\label{12.48a}
[A\circ B,C]_- &= A\circ[B,C]_\pm \mp [A,C]_\pm\circ B
\\				\label{12.48b}
[A\circ B,C]_+ &= A\circ[B,C]_\pm \mp [A,C]_\mp\circ B
	\end{align}
	\end{subequations}
from which, after simple but tedious algebraic calculations, follows
	\begin{equation}	\label{12.49}
\int_\sigma \bigl[
	\varphi_{i_1}[\sigma;x]\circ\dots\circ\varphi_{i_n}[\sigma;x]
	,
	\pi^{j\mu}[\sigma;y]
\bigr]_{\pm} \Id\sigma_\mu(y)
=
\ih c  \frac{\pd}{\pd\varphi_j[\sigma;x]}
\bigl(
	\varphi_{i_1}[\sigma;x]\circ\dots\circ\varphi_{i_n}[\sigma;x]
\bigr)
	\end{equation}
where $n\in\field[N]$ and the number of fermion operators/functionals, if
any, between $\varphi_{i_1},\ \dots,\ \varphi_{i_n}$ is \emph{even}.%
\footnote{~%
Notice, for the purposes of this paper, one can put, by definition,
\(
\frac{\pd}{\pd\varphi_{i_k}}
	( \varphi_{i_1}\circ\dots\circ\varphi_{i_n} )
=
(\pm1)^{k+1}
\varphi_{i_1}\circ\dots\circ
\varphi_{i_{k-1}}\circ\varphi_{i_{k+1}}
\circ\dots \circ\varphi_{i_n} ,
\)
$k=1,\dots,n$, if all fields are boson (sign plus) or fermion (sign minus)
ones. For some details regarding derivatives with respect to non\ndash
commuting variables, see~\cite{bp-QFT-action-principle}.%
}
Now the vanishment of the r.h.s.\ of~\eref{12.45} is a trivial corollary of
our supposition that
$\ope[\prime]{H}=-\ope[\prime\mspace{-3mu}]{L}$ or
$\tope[\prime]{H}=-\tope[\prime\mspace{-3mu}]{L}$
is a polynomial or convergent power series in the field operators/functionals
any term of which may contain only even number of fermion field components, if
any. \hfill Q.E.D.

	The \emph{Heisenberg equations of motion} for quantum fields in the
interaction picture (for arbitrary coupling) are an almost trivial corollary
of~\eref{12.38} with $\tope{A}(x)=\tope{\varphi}_i(x)$ and their
form~\eref{6.8} in Heisenberg picture:
	\begin{equation}	\label{12.50}
\ih\frac{\pd \varphi_i[\sigma;x] }{\pd x^\mu}
=
\bigl[ \varphi_i[\sigma;x] , \ope{P}_\mu[\sigma]
\bigr]_{\_}
+
\frac{1}{c} \Bigl[\ope[\prime]{H}[\sigma;x] ,
	\int_\sigma \varphi_i[\sigma;y] \Id\sigma_\mu(y) \Bigr]_{\_} .
	\end{equation}
Here
\(
\ope{P}_\mu[\sigma]
= \ope{U}[\sigma,\sigma_0]\circ \tope{P}_\mu
	\circ \ope{U}^{-1}[\sigma,\sigma_0]
\)
is the momentum operator in interaction picture.
In the particular case of nonderivative coupling, the second term in the
r.h.s.\ of~\eref{12.50} vanishes as a consequence of the supposed structure
of $\ope[\prime]{H}=-\ope[\prime\mspace{-3mu}]{L}$, \eref{12.48}, and the
(anti)commutation relations (see~\eref{4.5a})
	\begin{equation}	\label{12.51}
\bigl[ \varphi_i[\sigma;x] ,  \varphi_j[\sigma;y] \bigr]_{\pm} = 0
\qquad
(x-y)^2<0
\text{~~$x,y\in\sigma$, $x\not=y$} .
	\end{equation}
Hence, in view of~\eref{12.7}, equation~\eref{12.50} reduces to
	\begin{equation*}
\ih\frac{\pd \varphi_i[\sigma;x]  }{\pd x^\mu}
=
\bigl[ \varphi_i[\sigma;x] , \ope[0]{P}_\mu [\sigma]
\bigr]_{\_}
+
\delta_\mu^0
\Bigl[ \varphi_i[\sigma;x] ,
	\ope[\prime]{P}_0[\sigma]
\Bigr]_{\_} .
	\end{equation*}
Inserting here
\[
\ope[\prime]{P}_0[\sigma]
=
\frac{1}{c} \int_\sigma \Sprindex[{\ope[\prime]{T}}]{0}{\nu}[\sigma;y]
	\Id\sigma_\nu(y)
=
\frac{1}{c} \int_\sigma
	(-\Sprindex[\eta]{0}{\nu}\ope[\prime\mspace{-3mu}]{L}[\sigma;y])
	\Id\sigma_\nu(y)
=
\frac{1}{c} \int_\sigma \ope[\prime]{H}[\sigma;x]
	\Id\sigma_0(x)
\]
(see~\eref{12.3}--\eref{12.8}), we see that the second term in the last
equation vanishes by the same reasons as the one in~\eref{12.50}.
Consequently, as it should
be~\cite{Roman-QFT,Bogolyubov&Shirkov,Bjorken&Drell}, the Heisenberg
equations of motion in the interaction picture are
	\begin{equation}	\label{12.52}
\bigl[ \varphi_i[\sigma;x] , \ope[0]{P}_\mu [\sigma] \bigr]_{\_}
=
\ih\frac{\pd \varphi_i[\sigma;x]  }{\pd x^\mu}
	\end{equation}
provided the interaction Hamiltonian or Lagrangian contains nonderivative
coupling and is of the type specified above.%
\footnote{~%
The presented in~\cite[pp.~161--162]{Roman-QFT} derivation of~\eref{12.52}
starts from the equality
\(
\frac{\pd\varphi_i[\sigma;x]}{\pd x^\mu}
=
\ope{U}[\sigma,\sigma_0]
\circ \frac{\pd\tope{\varphi}_i[\sigma;x]}{\pd x^\mu} \circ
\ope{U}^\dag[\sigma_0,\sigma]
\)
which is not proved in \emph{loc.\ cit.}\ but it holds in the nonderivative
coupling case due to the vanishment of the second term in~\eref{12.50} in
this special situation.%
}

	Equation~\eref{12.52} agrees with~\eref{12.46}, both expressing the
fact that in the interaction picture the fields are solutions of the
corresponding \emph{free} equations. In particular, this entails the
assertion that all (anti)commutation relations for interacting fields in the
interaction picture are the same as for the corresponding free fields.

	A final remark at the end of this section. The actual computation
of the functional $\ope{U}[\sigma,\sigma_0]$ is a difficult task. The most
widely applied method for the purpose is the perturbation one. Its essence
is to rewrite the Tomonaga\ndash Schwinger equation~\eref{12.16} as a
Volterra integral equation, \viz
	\begin{equation}	\label{12.53}
\ope{U}[\sigma,\sigma_0]
=
\id_\Hil
+
\frac{1}{\ih c}
\int_{\sigma_0}^\sigma \ope[\prime]{H}[\sigma';x']
					\circ  \ope{U}[\sigma',\sigma_0]
\Id x' ,
	\end{equation}
where the integral is over all surfaces $\sigma'$ `between' $\sigma_0$ and
$\sigma$ from the foliation $\Sigma$ mentioned earlier, and, then to
solve~\eref{12.53} by successive iterations, starting from the initial value
 $\ope{U}^{(0)}[\sigma,\sigma_0]=\id_\Hil$. The result is the so\ndash called
T\ndash exponent (or P\ndash exponent, or chronological exponent)
	\begin{equation}	\label{12.54}
\ope{U}[\sigma,\sigma_0]
=
\Texp\Bigl(
\frac{1}{\ih c}
\int_{\sigma_0}^\sigma \ope[\prime]{H}[\sigma';x'] \Id x'
\Bigr)
	\end{equation}
on the successive approximations to which  are based the perturbation theory
and Feynman graph/diagram techniques.



\section
[Interaction picture. II.\ Time-dependent formulation]
{Interaction picture. II.\ Time-dependent formulation}
\label{Sect4}

	In this section is given a very concise presentation of the
time-dependent, non-covariant, treatment of the interaction picture. It is
\emph{implicitly} covariant and its comparison with the covariant one from
section~\ref{Sect3} shows what a big price is paid for the \emph{explicit}
covariance.

	Denote by $x_t$ a point in $\base$ such that $x^0=ct$ (in some Lorentz
frame of reference). The transition from Heisenberg picture to the
interaction one is provided by the following canonical transformations
(cf.~\eref{12.18})
	\begin{subequations}	\label{12.55}
	\begin{align}		\label{12.55a}
\tope{X}\mapsto \ope{X}(t)
	&:= \ope{U}(t,t_0) (\tope{X})
\\				\label{12.55b}
\tope{A}(x_t)\mapsto \ope{A}(x_t)
	&:= \ope{U}(t,t_0) \circ \tope{A}(x_t)\circ
	    \ope{U}^{-1}(t,t_0)
	\end{align}
	\end{subequations}
where $\tope{A}(x_t)\colon\Hil\to\Hil$, in particular
$\tope{A}(x_t)=\tope{\varphi}_i(x_t)$, and the unitary operator
$\tope{U}(t,t_0)\colon\Hil\to\Hil$,
$\tope{U}^\dag(t,t_0)=\tope{U}^{-1}(t_0,t)$, is the unique solution of the
initial\ndash value problem%
\footnote{~%
We distinguish $\tope{U}(t,t_0)$ from $\tope{U}[\sigma,\sigma_0]$ by using
parentheses and brackets, respectively, to denote their arguments.%
}
	\begin{align}		\label{12.56}
  \ih \frac{\pd \ope{U}(t,t_0)}{\pd t}
& =
\ope[\prime]{H}(t)\circ \ope{U}(t,t_0)
\\				\label{12.57}
\ope{U}(t_0,t_0) &= \id_\Hil
	\end{align}
in which
	\begin{equation}	\label{12.58}
\ope[\prime]{H}(t)
:=
\int \ope[\prime]{H}(x_t) \Id^3\bs x
=
\ope{U}(t,t_0) \circ \tope{H}(x_t)\circ \ope{U}^{-1}(t,t_0)
	\end{equation}
with
	\begin{equation}	\label{12.58new}
\tope[\prime]{H}(t)
:=
\int \tope[\prime]{H}(x_t) \Id^3\bs x
=
\int \tope[\prime]{H}(ct,\bs x) \Id^3\bs x
	\end{equation}
Thus $\ope{U}(t,t_0)$ is the solution of the equation
	\begin{align}	\label{12.58new1}
\ih\frac{\pd\ope{U}(t,t_0)}{\pd t}
&=
\ope{U}(t,t_0)\circ \tope[\prime]{H}(t)
\intertext{or}		\label{12.58new2}
\ih\frac{\pd\ope{U}^{-1}(t,t_0)}{\pd t}
&=
- \tope[\prime]{H}(t) \circ \ope{U}^{-1}(t,t_0)
	\end{align}
under the initial condition~\eref{12.57}. So, the $\ope{U}$-operator
$\ope{U}^{-1}(t,t_0)$ is given by the well\ndash known chronological exponent
	\begin{multline}	\label{12.58new3}
\ope{U}^{-1}(t,t_0)
=
\Texp\Bigl(
- \frac{1}{\ih}\int\limits_{t_0}^{t} \tope[\prime]{H}(t) \Id t
\Bigr)
\\
=
\id_\Hil
+
\sum_{n=1}^{\infty} \frac{(-1)^n}{(\ih)^n}
\int\limits_{t_0}^{t}\Id t_1
\int\limits_{t_0}^{t_1}\Id t_2
\dots
\int\limits_{t_0}^{t_{n-1}}\Id t_n
\tope[\prime]{H}(t_1) \circ\dots\circ \tope[\prime]{H}(t_n)
	\end{multline}
or, if $\ope[\prime]{H}(t)$ is known, by
	\begin{multline}	\label{12.59}
\ope{U}(t,t_0)
=
\Texp\Bigl(
\frac{1}{\ih}\int\limits_{t_0}^{t} \ope[\prime]{H}(t) \Id t
\Bigr)
\\
=
\id_\Hil
+
\sum_{n=1}^{\infty} \frac{1}{(\ih)^n}
\int\limits_{t_0}^{t}\Id t_1
\int\limits_{t_0}^{t_1}\Id t_2
\dots
\int\limits_{t_0}^{t_{n-1}}\Id t_n
\ope[\prime]{H}(t_1) \circ\dots\circ \ope[\prime]{H}(t_n) .
	\end{multline}
Notice, if the values of the interaction Hamiltonians at arbitrary moments
commute, \ie if
	\begin{equation}	\label{12.59new2}
[ \tope[\prime]{H}(t) , \tope[\prime]{H}(t') ]_{\_} = 0
	\end{equation}
for any moments $t$ and $t'$, which is equivalent to
	\begin{equation}	\tag{\ref{12.59new2}$^{\prime}$}
				\label{12.59new2'}
[ \ope[\prime]{H}(t) , \ope[\prime]{H}(t') ]_{\_} = 0,
	\end{equation}
then~\eref{12.59} implies the commutativity of $\ope{U}(t,t_0)$  and
$\ope[\prime]{H}(t)$, \ie
	\begin{equation}	\label{12.59new3}
[ \ope{U}(t,t_0) , \ope[\prime]{H}(t) ]_{\_} = 0 ,
	\end{equation}
which, in its turn, entails
	\begin{equation}	\label{12.59new4}
\ope[\prime]{H}(t) = \tope[\prime]{H}(t)  .
	\end{equation}

	\begin{Rem}
				\label{Rem12.1}
	\begin{small}
Since initial-value problems like~\eref{12.56}--\eref{12.57} will be met and
further in this work, here is a concise summary of their theory. Suppose
$G(s)\colon\Hil\to\Hil$ is an operator depending continuously on a real
parameter $s$. For a fixed $s_0\in\field[R]$, the solution of
	\begin{equation}	\label{12.Texp1}
\frac{\pd Y(s,s_0)}{\pd s} = G(s)\circ Y(s,s_0)
\qquad
Y(s_0,s_0) = \id_\Hil
	\end{equation}
is called \emph{chronological exponent} (T-\emph{exponent}) of $s$ and
	\begin{align}	\label{12.Texp2}
Y(s,s_0)
& =
\Texp\Bigl( \int_{s_0}^{s} G(\tau) \Id\tau \Bigr)
\\ \notag
& :=
\id_\Hil
+ \sum_{n=1}^{\infty} 	\int_{s_0}^{s}\Id s_1 \int_{s_0}^{s_1}\Id s_2
			\cdots \int_{s_0}^{s_{n-1}}\Id s_n
	G(s_1)\circ\dots\circ G(s_n)
\\ \notag
& =
\id_\Hil
+ \sum_{n=1}^{\infty} 	\int_{s_0}^{s}\Id s_1 \int_{s_0}^{s}\Id s_2
			\cdots \int_{s_0}^{s}\Id s_n
	\mathrm{T} \bigl( G(s_1)\circ\dots\circ G(s_n) \bigr) ,
	\end{align}
where
\(
\mathrm{T} \bigl( G(s_1)\circ\dots\circ G(s_n) \bigr)
:=
G(s_{i_1})\circ\dots\circ G(s_{i_n}) \bigr)
\)
with $i_1,\dots,i_n$ being a permutation of $1,\dots,n$ such that
 $s_1\ge s_2\ge \dots \ge s_n$.
	For any $s,s_0,s_1\in\field[R]$, we have
	\begin{gather}	\label{12.Texp3}
(Y(s,s_0))^{-1} = Y(s_0,s)
\\			\label{12.Text4}
Y(s,s_1)\circ Y(s_1,s_0) = Y(s,s_0).
	\end{gather}
If $G$ is anti-Hermitian,
	\begin{equation}	\label{12.Texp5}
(G(s) )^\dag = - G(s) ,
	\end{equation}
then the chronological exponent is unitary, \viz
	\begin{equation}	\label{12.Texp6}
(Y(s,s_0))^\dag = (Y(s,s_0))^{-1} =: Y^{-1}(s,s_0).
	\end{equation}
Useful corollaries from~\eref{12.Texp1} and~\eref{12.Texp3} are:
	\begin{gather}	\label{12.Texp7}
\frac{\pd Y^{-1}(s,s_0)}{\pd s} = - Y^{-1}(s,s_0) \circ  G(s)
\\			\label{12.Text8}
\frac{\pd Y(s_0,s)}{\pd s} = - Y(s_0,s) \circ  G(s) .
	\end{gather}
If $G_0\colon\Hil\to\Hil$ is a constant operator and
$f\colon\field[R]\to\field[C]$ is an integrable function, then
	\begin{gather}
				\label{12.Texp9}
\Texp\Bigl( \int_{s_0}^{s} G_0 \Id\tau \Bigr)
=\exp ((s-s_0)G_0)
\\				\label{12.Texp10}
\Texp\Bigl( \int_{s_0}^{s} f(\tau) \id_\Hil \Id\tau \Bigr)
=
\exp\Bigl( \int_{s_0}^{s} f(\tau)  \Id\tau \Bigr) \id_\Hil .
\\\intertext{If it happens that $G(s)$ commutes with $Y(s,s_0)$,}
				\label{12.Texp11}
[G(s),Y(s,s_0)]_{\_} = 0 ,
\\\intertext{a sufficient condition for which is}
				\label{12.Texp12}
[G(s),G(s')]_{\_} = 0
\\\intertext{for any $s,s'\in\field[R]$, then $Y(s,s_0)$ is also a solution
of}
				\label{12.Texp13}
\frac{\pd Y(s,s_0)}{\pd s} = Y(s,s_0)\circ G(s)
\qquad
Y(s_0,s_0) = \id_\Hil .
	\end{gather}
At last, the above-considered interaction picture corresponds to the case
$G(t)=\iih\ope[\prime]{H}(t)$.\vspace{1.7ex}
	\end{small}
	\end{Rem}

	In particular, the field operators $\tope{\varphi}_i(x_t)$ transform
into
	\begin{equation}	\label{12.60}
\varphi_i(x_t)
=
\ope{U}(t,t_0) \circ \tope{\varphi}_i(x_t)\circ  \ope{U}^{-1}(t,t_0) .
	\end{equation}

	From~\eref{12.55a}, \eref{12.56}, and \eref{12.57},
the equation of motion for the state vectors in the (time\ndash dependent)
interaction picture immediately follow  (cf.~\eref{12.36} and~\eref{12.37}):
	\begin{align}	\label{12.61}
\ih \frac{\od\ope{X}(t)}{\od x}
&=
\ope[\prime]{H}(t) (\ope{X}(t))
\\			\label{12.62}
\ope{X}(t_0) &= \tope{X} .
	\end{align}

	To derive the equations of motion for the field operators
$\varphi_i(x_t)$, we shall use the equality (cf.~\eref{12.38})
	\begin{equation}	\label{12.63}
\ih \frac{\pd\ope{A}(x_t)}{\pd t}
=
\ope{U}(t,t_0) \circ \ih \frac{\pd\tope{A}(x_t)}{\pd t}\circ
\ope{U}^\dag(t_0,t)
+
[ \ope[\prime]{H}(t),\ope{A}(x_t) ]_{\_}
	\end{equation}
which is a simple consequence of~\eref{12.55b} and~\eref{12.56}. The
particular choices
$\tope{A}(x_t)=\tope{\varphi}_i(x_t)$ and
\(
\tope{A}(x_t)
= \frac{1}{c}\tope{\pi}^{i0}(x_t)
= \frac{\pd\tope{L}}{\pd(\pd_t\tope{\varphi}_i(x_t) )}
\)
reduce~\eref{12.63} respectively to%
\footnote{~%
Here and henceforth the operator $\pd_t:=\frac{\pd}{\pd t}$ denotes the
partial derivative with respect to the time coordinate $t=x^0/c$.%
}
	\begin{align}	\label{12.64}
\ih \frac{\pd\ope{\varphi}_i(x_t)}{\pd t}
& =
\ope{U}(t,t_0) \circ \ih \frac{\pd \tope{\varphi}_i(x_t)}{\pd t}\circ
\ope{U}^\dag(t_0,t)
+
[ \ope[\prime]{H}(t),\ope{\varphi}_i(x_t) ]_{\_}
\\			\label{12.65}
\ih \frac{\pd \frac{\pi^{i0}(x_t)}{c} }{\pd t}
& =
\ope{U}(t,t_0) \circ \ih
\frac{\pd}{\pd t}\Bigl(
		\frac{\pd\tope{L}}{\pd(\pd_t\tope{\varphi}_i(x_t) )}
		 \Bigr)
\circ  \ope{U}^\dag(t_0,t)
+
\Bigl[ \ope[\prime]{H}(t),\frac{\pi^{i0}(x_t)}{c} \Bigr]_{\_} ,
	\end{align}
where, in view of~\eref{12.64},
\footnote{~%
When we differentiate the Lagrangian/Hamiltonian with respect to
$\pd_t\varphi_i(x_t)$, we mean that this symbol denotes the corresponding
argument of the Lagrangian/Hamiltonian and \emph{not} the values of
$\pd_t\varphi_i\equiv\frac{\pd\varphi_i}{\pd t}$ at a point $x_t$, \ie not
$\pd_t\varphi_i|_{x_t}=\frac{\pd\varphi_i(x_t)}{\pd t}$.%
}
	\begin{multline*}
\frac{1}{c}\ope{\pi}^{i0}(x_t)
:=
\ope{U}(t,t_0) \circ
\frac{\pd\tope{L}}{\pd(\pd_t\tope{\varphi}_i(x_t) )}
\circ  \ope{U}^\dag(t_0,t)
=
\ope{U}(t,t_0) \circ
\frac{\pd\tope{L}}{\pd(\pd_t\ope{\varphi}_j(x_t) )}
\\
\circ
       \ope{U}^\dag(t_0,t) \circ
       \ope{U}(t,t_0) \circ
\frac{\pd(\pd_t\ope{\varphi}_j(x_t) )}
     {\pd(\pd_t\tope{\varphi}_i(x_t) )}
\circ  \ope{U}^\dag(t_0,t)
=
\frac{\pd\ope{L}}{\pd(\pd_t\ope{\varphi}_i(x_t) )}
+
\ope{G}^i
	\end{multline*}
with 
	\begin{equation}	\label{12.66}
\ope{G}^i :=
\frac{1}{\ih}
\frac{\pd\ope{L}}{\pd(\pd_t\ope{\varphi}_j(x_t) )} \circ
\Bigl[
\frac{\pd\ope[\prime]{H}(t)}{\pd(\pd_t\tope{\varphi}_i(x_t) )}
,
\ope{\varphi}_j(x_t)
\Bigr]_{\_}
= 0 .
	\end{equation}
The last equality is a consequent of the fact that $\ope[\prime]{H}(t)$, in
view of~\eref{12.58} and~\eref{12.58new}, is operator\ndash valued
functional, not a function, of $\varphi_i$ and $\pd_\mu\varphi_i$ which
implies
	\begin{equation*}
\frac{\pd \ope[\prime]{H}(t)}{\pd \tope{\varphi}_i(x_t)} \equiv 0
\qquad
\frac{\pd \ope[\prime]{H}(t)}{\pd (\pd_\mu\tope{\varphi}_i(x_t))} \equiv 0 .
	\end{equation*}
Useful corollaries from these identities,~\eref{12.60},~\eref{12.64},
and~\eref{12.65} are:
	\begin{equation}	\label{12.66new}
\frac{\pd}{\pd \tope{\varphi}_i(x_t)}
=
\frac{\pd}{\pd \ope{\varphi}_i(x_t)}
\qquad
\frac{\pd}{\pd (\pd_\mu\tope{\varphi}_i(x_t))}
=
\frac{\pd}{\pd (\pd_\mu\ope{\varphi}_i(x_t))}
	\end{equation}

	Hence, due to~\eref{12.64},
	\begin{multline}	\label{12.67}
\ope{L}
=
\ope{L}\bigl( \varphi_i(x_t),\pd_t\varphi_j(x_t), \pd_a\varphi_j(x_t) \bigr)
\\ :=
\ope{U}(t,t_0) \circ
\tope{L}\bigl( \tope{\varphi}_i(x_t), \pd_t\tope{\varphi}_j(x_t),
	  \pd_a\tope{\varphi}_j(x_t) \bigr)
\circ  \ope{U}^\dag(t_0,t)
\\ =
\tope{L}\bigl( \varphi_i(x_t),
	\ope{U}(t,t_0) \circ
	\pd_t\tope{\varphi}_j(x_t) \circ
	\ope{U}^\dag(t_0,t)
	,
	\pd_a\varphi_j(x_t) \bigr)
\\=
\tope{L}\bigl( \varphi_i(x_t),
	\pd_t\varphi_j(x_t)
	- \frac{1}{\ih}
	[ \ope[\prime]{H}(t),\ope{\varphi}_j(x_t) ]_{\_}
	,
	\pd_a\varphi_j(x_t) \bigr)
	\end{multline}
with $a=1,2,3$. Substituting in~\eref{12.65} the above expression for
$\frac{1}{c}\ope{\pi}^{i0}(x_t)$ and
	\begin{equation*}
\frac{\pd}{\pd t}\Bigl(
		\frac{\pd\tope{L}}{\pd(\pd_t\tope{\varphi}_i(x_t) )}
		 \Bigr)
=
\frac{\pd\tope{L}}{\pd \tope{\varphi}_i(x_t)}
-
\sum_{a=1}^{3}
\frac{\pd}{\pd x^a}\Bigl(
		\frac{\pd\tope{L}}{\pd(\pd_a\tope{\varphi}_i(x_t) )}
		 \Bigr) ,
	\end{equation*}
which follows from the Euler-Lagrange equation~\eref{4.1},
and using~\eref{12.67}, we obtain (cf.~\eref{12.43})
	\begin{equation}	\label{12.68}
\frac{\pd\ope{L}}{\pd \ope{\varphi}_i(x_t)}
-
\frac{\pd}{\pd x^\mu}\Bigl(
		\frac{\pd\ope{L}}{\pd(\pd_\mu\ope{\varphi}_i(x_t) )}
		 \Bigr)
=
- \frac{1}{\ih}
\Bigl[
\ope[\prime]{H}(t) ,
\frac{\pd\ope{L}}{\pd(\pd_t\ope{\varphi}_i(x_t) )}
\Bigr]_{\_} .
	\end{equation}
These are the \emph{Euler-Lagrange equations of motion} for the quantum
fields in the (time\ndash dependent) interaction picture.

	In a case of nonderivative coupling, we have
$\ope{L}=\ope[0\mspace{-3mu}]{L}+\ope[\prime\mspace{-3mu}]{L}$, in which
$\ope[\prime\mspace{-3mu}]{L}=-\ope[\prime]{H}$ is independent of
$\pd_\mu\varphi_i(x_t)$, so that~\eref{12.68} reduces to
	\begin{equation*}
\frac{\pd\ope[0\mspace{-3mu}]{L}}{\pd \ope{\varphi}_i(x_t)}
-
\frac{\pd}{\pd x^\mu}\Bigl(
	\frac{\pd\ope[0\mspace{-3mu}]{L}}{\pd(\pd_\mu\ope{\varphi}_i(x_t) )}
	 \Bigr)
=
\frac{\pd \ope[\prime]{H}(t)}{\pd\varphi_i(x_t)}
- \frac{1}{\ih}
\Bigl[
	\ope[\prime]{H}(t) ,
	\frac{\pd\ope[0\mspace{-3mu}]{L}}{\pd(\pd_t\ope{\varphi}_i(x_t) )}
\Bigr]_{\_} .
	\end{equation*}
The r.h.s.\ of the last equation identically vanishes as a result of the
equal-time canonical (anti)communion relations ~\eref{4.5c}
$\big($in our case
\(
\pi^{i}(x^0,\bs x)=\pi^i(x_t)
=
\frac{\pd\ope[0\mspace{-3mu}]{L}}{\pd(\pd_t\ope{\varphi}_i(x_t) )}
\)$\big)$
and arguments similar to the ones leading from~\eref{12.45} to~\eref{12.46}.
Therefore, in a case of nonderivative coupling, the Euler\ndash Lagrange
equations of motion in the (time\ndash dependent) interaction picture are
	\begin{equation}	\label{12.70}
\frac{\pd\ope[0\mspace{-3mu}]{L}}{\pd \ope{\varphi}_i(x_t)}
-
\frac{\pd}{\pd x^\mu}\Bigl(
	\frac{\pd\ope[0\mspace{-3mu}]{L}}{\pd(\pd_\mu\ope{\varphi}_i(x_t) )}
	 \Bigr)
 = 0
	\end{equation}
and, as one can expect, coincide with the \emph{free} equations for the
\emph{non-free} fields.

	The derivation of Heisenberg equation of motion in the time-dependent
interaction picture is completely trivial. Recalling that in the Heisenberg
picture they are given by~\eref{6.8}, after the canonical
transformation~\eref{12.55} they take the form (cf.~\eref{12.50})
	\begin{subequations}	\label{12.71}
	\begin{align}		\label{12.71a}
\ih\frac{\pd}{\pd t} \varphi_i(x_t)
& =
[\varphi_i(x_t), c\ope{P}_0 - \ope[\prime]{H}(t) ]_{\_}
=
[\varphi_i(x_t), c\ope[0]{P}_0 ]_{\_}
\\				\label{12.71b}
\ih\frac{\pd}{\pd x^a} \varphi_i(x_t)
& =
[\varphi_i(x_t), \ope{P}_a ]_{\_}
\qquad
a=1,2,3 ,
	\end{align}
	\end{subequations}
where~\eref{12.64} and $\ope{P}_0=\frac{1}{c}\ope{H}(t)$ were
taken into account for the derivation of~\eref{12.71a} which, regardless of
the coupling, derivative or nonderivative, always has a form of a free
equation. Of course, for a nonderivative coupling these equations take the
free form
	\begin{equation}	\label{12.72}
\ih\frac{\pd}{\pd x^\mu} \varphi_i(x_t)
=
[\varphi_i(x_t), \ope[0]{P}_\mu ]_{\_}
	\end{equation}
as a result of~\eref{12.7}.


\section {Schr\"odinger picture}
\label{Sect5}

	The connection between Heisenberg and Schr\"odinger pictures in
quantum field theory is similar to the one in (nonrelativistic) quantum
mechanics~\cite{Messiah-QM,Dirac-PQM,Fock-FQM} with the simplification that
in field theory the (total) Hamiltonian is a constant in time operator as
closed systems (with conserve 4\ndash momentum) are considered. Since in the
general formalism this simplification is not quite essential, we shall
neglect it and, respectively, the Hamiltonian will be written with a time
argument $t$. The idea of Schr\"odinger picture is the time dependence of
the field operators  and observables constructed from them (in, e.g.,
Heisenberg picture) to be transferred entirely on the state vectors, \ie the
former ones should become time\ndash independent, while the latter ones
become time\ndash dependent.

	Suppose, we have a system of quantum fields with 4-momentum
$\tope{P}_\mu$ and Hamiltonian
$\tope{H}(t)=c\tope{P}_0$, both given in the Heisenberg picture
(in which all quantities are labeled by tilde above their kernel symbol).
The transition from the Heisenberg picture to
Schr\"odinger one is performed in the same way as from Heisenberg picture to
the time\ndash dependent interaction one, as described in
Sect.~\ref{Sect4}, with the only difference that the
\emph{interaction Hamiltonian $\tope[\prime]{H}(t)$ must be replaced by the
total Hamiltonian}
$\tope{H}(t)=\tope[0]{H}(t)+\tope[\prime]{H}(t)$
(which is time\ndash independent for closed systems). Therefore, the
mappings~\eref{12.55} realize the transition (from Heisenberg) to
Schr\"odinger picture if the $\ope{U}$\ndash operator $\ope{U}(t,t_0)$ is the
unique solution of the equation (cf.~\eref{12.56})
	\begin{equation}		\label{12.73}
\ih \frac{\pd \ope{U}(t,t_0)}{\pd t}
=
\ope{H}(t)\circ \ope{U}(t,t_0)
	\end{equation}
under the initial conditions~\eref{12.57}, $\ope{U}(t_0,t_0)=\id_\Hil$.
Here
	\begin{equation}	\label{12.74}
\ope{H}(t)
:=
\int \ope{H}(x_t) \Id^3\bs x
=
\ope{U}(t,t_0) \circ \tope{H}(x_t)\circ \ope{U}^{-1}(t,t_0)
	\end{equation}
with
	\begin{equation}	\label{12.75}
\tope{H}(t)
:=
\int \tope{H}(x_t) \Id^3\bs x
=
\int \tope{H}(ct,\bs x) \Id^3\bs x .
	\end{equation}
So, if $\tope{H}(t)$ is given, $\ope{U}(t,t_0)$ is a solution of
(cf.~\eref{12.58new1} and~\eref{12.58new2})
	\begin{align}	\label{12.76}
\ih\frac{\pd\ope{U}(t,t_0)}{\pd t}
=
\ope{U}(t,t_0)\circ \tope{H}(t)
\quad\text{or}\quad
\ih\frac{\pd\ope{U}^{-1}(t,t_0)}{\pd t}
=
- \tope{H}(t) \circ \ope{U}(t,t_0) .
	\end{align}
If we take into account that for a closed system $\tope{H}(t)$ is a constant
of motion, \ie $\frac{\pd\tope{H}(t)}{\pd t}=0$ or
$\tope{H}(t)=\tope{H}(t_0)\equiv\tope{H}$, then,
by~\eref{12.Texp9}--\eref{12.Texp13},
	\begin{align}	\label{12.76new}
\ope{H}(t) &= \tope{H}
\\			\label{12.76new1}
\ope{U}(t,t_0) &= \e^{\frac{1}{\ih}(t-t_0)\ope{H}}
	\end{align}
which considerably simplifies some calculations.

	In Schr\"odinger picture the state vectors, given via~\eref{12.55a}
with the operator $\ope{U}(t,t_0)$ defined above, are solutions of the
initial\ndash value problem (cf.~\eref{12.61} and~\eref{12.62})
	\begin{align}	\label{12.77}
\ih \frac{\od\ope{X}(t)}{\od x}
=
\ope{H}(t) (\ope{X}(t))
\qquad
\ope{X}(t_0) = \tope{X} .
	\end{align}

	The Euler-Lagrange equations for the field operators, given
via~\eref{12.60} with above $\ope{U}(t,t_0)$, are (cf.~\eref{12.68})
	\begin{equation}	\label{12.78}
\frac{\pd\ope{L}}{\pd \ope{\varphi}_i(x_t)}
-
\frac{\pd}{\pd x^\mu}\Bigl(
		\frac{\pd\ope{L}}{\pd(\pd_\mu\ope{\varphi}_i(x_t) )}
		 \Bigr)
=
- \frac{1}{\ih}
\Bigl[
\ope{H}(t) ,
\frac{\pd\ope{L}}{\pd(\pd_t\ope{\varphi}_i(x_t) )}
\Bigr]_{\_} .
	\end{equation}
Here,
	\begin{equation}	\label{12.78new}
\ope{L}
=
\tope{L}\bigl( \varphi_i(x_t),
	\pd_t\varphi_j(x_t)
	- \frac{1}{\ih}
	[ \ope{H}(t),\ope{\varphi}_j(x_t) ]_{\_}
	,
	\pd_a\varphi_j(x_t) \bigr) ,
	\end{equation}
where $a=1,2,3$, is the Lagrangian in Schr\"odinger picture and,
as in Sect.~\ref{Sect4}, in the partial derivative
$\frac{\pd}{\pd(\pd_t\varphi_j(x_t)}$ the expression $\pd_t\varphi_j(x_t)$
means the corresponding argument of the Lagrangian/Hamiltonian, and
\emph{not} the value of $\pd_t\varphi_j$ at a point $x_t$, \ie not
$\frac{\pd\varphi_j(x_t)}{\pd t}=\pd_t\varphi_j|_{x_t}$ which identically
vanishes (in the Schr\"odinger picture) by the proved below
equation~\eref{12.81a}. (The last means that one, at first, has to perform
the differentiation relative to $\pd_t\varphi_j(x_t)$ and, then, to set this
argument to zero.)
Repeating the procedure leading from~\eref{12.68} to~\eref{12.70}, we see
that in Schr\"odinger picture the Euler\ndash Lagrange equations for a
nonderivative coupling are (cf.~\eref{12.70})
	\begin{equation}	\label{12.80}
\frac{\pd\ope[0\mspace{-3mu}]{L}}{\pd \ope{\varphi}_i(x_t)}
-
\frac{\pd}{\pd x^\mu}\Bigl(
	\frac{\pd\ope[0\mspace{-3mu}]{L}}{\pd(\pd_\mu\ope{\varphi}_i(x_t) )}
	 \Bigr)
 =
- \frac{1}{\ih}
\Bigl[
	\ope[0]{H}(t) ,
	\frac{\pd\ope[0\mspace{-3mu}]{L}}{\pd(\pd_t\ope{\varphi}_i(x_t) )}
\Bigr]_{\_} .
	\end{equation}
A few lines below, it will be proved that the terms, containing time
derivatives in~\eref{12.78} and~\eref{12.80}, vanish and,
consequently, equations~\eref{12.78} and~\eref{12.80}  are equivalent
to~\eref{12.81new} and~\eref{12.81new1}, respectively.

	The Heisenberg equations of motion for $\varphi_i(x_t)$ in the
Schr\"odinger picture are (cf.~\eref{12.71})
	\begin{subequations}	\label{12.81}
	\begin{align}		\label{12.81a}
\ih\frac{\pd}{\pd t} \varphi_i(x_t)
& = 0
\\				\label{12.81b}
\ih\frac{\pd}{\pd x^a} \varphi_i(x_t)
& =
[\varphi_i(x_t), \ope{P}_a ]_{\_}
\qquad
a=1,2,3
	\end{align}
	\end{subequations}
and are obtained by the same method as~\eref{12.71} with the only difference
that $\ope{H}(t) = c\ope{P}_0$ should be used instead of
$\ope[\prime]{H}(t)$. In the nonderivative coupling case,~\eref{12.81b}
transforms into the same equality with
 $\ope[0]{P}_a$ for  $\ope{P}_a$, $a=1,2,3$,
but~\eref{12.81a} remains unchanged.

	Equation~\eref{12.81a} shows that, as we stated at the beginning of
the present section, in Schr\"odinger picture the field operators are
time\ndash independent. Obviously, the same is true for operators constructed
from them and their spacial derivatives (of finite order).

	These observations imply an important corollary: since the
Lagrangians and Hamiltonians are supposed to be constructed from the field
operators and their first partial derivatives as polynomials or convergent
power series, the time derivatives of a Lagrangian/Hamiltonian or some its
partial derivatives (with respect to coordinates and/or field operators or
their first partial derivatives) identically vanish in the Schr\"odinger
picture. In particular, the terms containing time derivatives in~\eref{12.78}
vanish. This proves that~\eref{12.78} is equivalent to
	\begin{equation}	\label{12.81new}
\frac{\pd\ope{L}}{\pd \ope{\varphi}_i(x_t)}
-
\sum_{a=1}^{3}\frac{\pd}{\pd x^a}\Bigl(
		\frac{\pd\ope{L}}{\pd(\pd_a\ope{\varphi}_i(x_t) )}
		 \Bigr)
=
- \frac{1}{\ih}
\Bigl[
\ope{H}(t) ,
\frac{\pd\ope{L}}{\pd(\pd_t\ope{\varphi}_i(x_t) )} 
\Bigr]_{\_}
	\end{equation}
which, in view of~\eref{12.80}, in the nonderivative coupling case reduces
to
	\begin{equation}	\label{12.81new1}
\frac{\pd\ope[0\mspace{-3mu}]{L}}{\pd \ope{\varphi}_i(x_t)}
-
\sum_{a=1}^{3}\frac{\pd}{\pd x^a}\Bigl(
	\frac{\pd\ope[0\mspace{-3mu}]{L}}{\pd(\pd_a\ope{\varphi}_i(x_t) )}
	 \Bigr)
 =
- \frac{1}{\ih}
\Bigl[
	\ope[0]{H}(t) ,
	\frac{\pd\ope[0\mspace{-3mu}]{L}}{\pd(\pd_t\ope{\varphi}_i(x_t) )}
\Bigr]_{\_} .
	\end{equation}

	The above presentation of the Schr\"odinger picture can be called
time\ndash dependent due to its explicit dependence on the time coordinates.
However, this exposition  turns to be \emph{implicitly covariant} and it
admits an explicit covariant formulation which is completely similar to the
one of the interaction picture in Sect.~\ref{Sect3}. The covariant
formulation of the Schr\"odinger picture can be obtained from the one of
interaction picture, given in Sect.~\ref{Sect3}, by replacing the
interaction Hamiltonians $\tope[\prime]{H}(x)$ and $\ope[\prime]{H}[\sigma;x]$
by the corresponding total Hamiltonians
$\tope{H}(x)=\tope[0]{H}(x)+\tope[\prime]{H}(x)$ and
$\ope{H}[\sigma;x]=\ope[0]{H}[\sigma;x]+\ope[\prime]{H}[\sigma;x]$.
Below we give a concise sketch of this procedure.

	Suppose a unitary operator
$\ope{U}[\sigma,\sigma_0]\colon\Hil\to\Hil$ is defined as the solution of
	\begin{equation}	\label{12.80new1}
\ih c \frac{\delta\ope{U}[\sigma,\sigma_0]}{\delta\sigma(x)}
=
\ope{H}[\sigma;x]\circ \ope{U}[\sigma,\sigma_0]
\quad
x\in\sigma
	\end{equation}
satisfying the initial condition
	\begin{equation}	\label{12.80new2}
\ope{U}[\sigma_0,\sigma_0] = \id_\Hil
	\end{equation}
where 
	\begin{equation}	\label{12.80new3}
\ope{H}[\sigma;x]
:= \ope{U}[\sigma,\sigma_0] \circ \tope{H}(x) \circ
   	    \ope{U}^{-1}[\sigma,\sigma_0]
\quad
x\in\sigma
	\end{equation}
with $\tope{H}(x)$ being the (total) Hamiltonian in Heisenberg picture.%
\footnote{~%
Actually $\ope{H}[\sigma;x]$ and $\tope{H}(x)=c\ope{H}_{0}$ do
not depend on $x$ by virtue of the conservation of 4\ndash momentum for a
closed (translation invariant) system. This simplification is not quite
essential for the following.%
}
The general transformations~\eref{12.18}, with above defined
`$\ope{U}$\ndash operator', give the transition from Heisenberg picture to
Schr\"odinger one in covariant formulation.

	The evolution of the state functionals~\eref{12.18a} is governed by
the initial\ndash value problem ($x\in\sigma$)
	\begin{align}	\label{12.80new4}
\ih c \frac{\delta\ope{X}[\sigma]}{\delta\sigma(x)}
=
\ope{H}[\sigma;x] (\ope{X}[\sigma])
\qquad
\ope{X}[\sigma]|_{\sigma=\sigma_0} = \ope{X}[\sigma_0] = \tope{X} ,
	\end{align}
while the field functionals~\eref{12.20} are solutions of the Euler\ndash
Lagrange equations~\eref{12.43} in which
	\begin{multline}	\label{12.80new5}
\ope{G}^{i\mu}[\sigma;x]
=
\ope{G}^{i\mu}\bigl( \varphi_i[\sigma;x],\pd_\mu\varphi_j[\sigma;x] \bigr)
\\
:=
\frac{1}{\ih c}
\frac{\pd\ope{L}}{\pd( \pd_\nu\ope{\varphi}_j[\sigma;x] )}
	\circ
	\Bigl[
    \frac{\pd\ope{H}[\sigma;x]}  {\pd( \pd_\mu\tope{\varphi}_i(x) )}
	,
    \int_\sigma \varphi_j[\sigma;y] \Id\sigma_\nu(y)
	\Bigr]_{\_}
\Bigr\} .
	\end{multline}

	At last, the Heisenberg equations of motion are (cf.~\eref{12.50})
	\begin{equation}	\label{12.80new6}
\ih\frac{\pd \varphi_i[\sigma;x] }{\pd x^\mu}
=
\bigl[ \varphi_i[\sigma;x] , \ope{P}_\mu[\sigma]
\bigr]_{\_}
+
\frac{1}{c} \Bigl[\ope{H}[\sigma;x] ,
	\int_\sigma \varphi_i[\sigma;y] \Id\sigma_\mu(y) \Bigr]_{\_} .
	\end{equation}
From where, for $\mu=0$, we get
	\begin{equation}	\label{12.80new7}
\ih\frac{\pd \varphi_i[\sigma;x] }{\pd x^0} = 0
	\end{equation}
due to $\ope{H}[\sigma;x]=c\ope{P}_{0}[\sigma]$.


\section
[Links between different time-dependent pictures of motion]
{Links between different time-dependent pictures of motion}
\label{Sect6}

	In the present section, we briefly summarize the connections between
different time\ndash dependent pictures (representations) of motion in
Lagrangian quantum field theory.

	Let the index $\omega$ labels a given picture of motion; in
particular, it can take the values $\mathrm{H}$, $\mathrm{S}$, and
$\mathrm{I}$ for, respectively, Heisenberg, Schr\"odinger, and Interaction
picture.

	Suppose, the \emph{time evolution} of a state vector
$\ope{X}^\omega(x_t)\in\Hil$ is described via an \emph{evolution operator}
$\ope{U}^\omega(t,t_0)$, \viz
	\begin{equation}	\label{12.82}
\ope{X}^\omega(x_t) = \ope{U}^\omega(t,t_0) (\ope{X}^\omega(x_{t_0}))
	\end{equation}
for any instants of time $t$ and $t_0$. The operator $\ope{U}^\omega(t,t_0)$
is, by definition, unitary and is defined as the unique solution of the
initial\ndash value problem
	\begin{align}	\label{12.83}
\frac{\pd\ope{U}^\omega(t,t_0)}{\pd t}
&=
\underline{\ope{H}}^\omega(t)\circ \ope{U}^\omega(t,t_0)
\\			\label{12.84}
\ope{U}^\omega(t_0,t_0)
&=
\id_\Hil
	\end{align}
where $\underline{\ope{H}}^\omega(t)$ is a given operator-valued function of
time, connected to the Hamiltonian in the  $\omega$\ndash picture and, hence,
it is an operator\ndash valued functional of the field operators and their
derivatives. Notice,~\eref{12.83} corresponds to the Schr\"odinger equation
	\begin{align}	\label{12.85}
\frac{\pd\ope{X}^\omega(x_t)}{\pd t}
=
\underline{\ope{H}}^\omega(t) (\ope{X}^\omega(x_t))
	\end{align}
for a state vector $\ope{X}^\omega(x_t)$.

	The transition from an $\omega$-picture to an $\omega'$-picture is
performed by means of an unitary  `$\ope{U}$\ndash operator'
	\begin{equation}	\label{12.85new}
\ope{U}^{\omega\to\omega'} (t,t_0)
:=
\ope{U}^{\omega'} (t,t_0) \circ (\ope{U}^{\omega} (t,t_0) )^{-1},
	\end{equation}
that is, we have
	\begin{subequations}	\label{12.86}
	\begin{align}	\label{12.86a}
\ope{X}^\omega(x_t)\mapsto \ope{X}^{\omega'}(x_t)
&:=
\ope{U}^{\omega\to\omega'} (t,t_0) \bigl( \ope{X}^\omega(x_t) \bigr)
\\			\label{12.86b}
\ope{A}^\omega(x_t)\mapsto \ope{A}^{\omega'}(x_t)
&:=
\ope{U}^{\omega\to\omega'} (t,t_0)
\circ \ope{A}^\omega(x_t) \circ (\ope{U}^{\omega\to\omega'} (t,t_0) )^{-1} .
	\end{align}
	\end{subequations}
The `$\ope{U}$\ndash operator' is a solution of the initial-value problem
	\begin{align}	\label{12.87}
\ih\frac{\pd\ope{U}^{\omega\to\omega'}(t,t_0)}{\pd t}
&=
\underline{\ope{H}}^{\omega'}(t) \circ \ope{U}^{\omega\to\omega'} (t,t_0)
-
\ope{U}^{\omega\to\omega'} (t,t_0) \circ \underline{\ope{H}}^{\omega}(t)
\\			\label{12.88}
\ope{U}^{\omega\to\omega'} (t_0,t_0)
&=
\id_\Hil
	\end{align}
which is a consequence of~\eref{12.83},~\eref{12.84}, and~\eref{12.85new}.

	In particular, if one stars from, e.g., the Schr\"odinger picture,
then  $\underline{\ope{H}}^{\mathrm{S}}=\ope{H}^{\mathrm{S}}$ is the
Hamiltonian in the Schr\"odinger picture, $\ope{U}^{\mathrm{S}}$ is a
solution of
	\begin{align}	\label{12.89}
\frac{\pd\ope{U}^{\mathrm{S}}(t,t_0)}{\pd t}
=
\ope{H}^{\mathrm{S}}(t)\circ \ope{U}^{\mathrm{S}}(t,t_0)
\qquad
\ope{U}^{\mathrm{S}}(t_0,t_0)
=
\id_\Hil
	\end{align}
and
	\begin{subequations}	\label{12.90}
	\begin{align}	\label{12.90a}
	\begin{split}
\underline{\ope{H}}^{\mathrm{H}}(t) = 0
\qquad
\ope{U}^{\mathrm{H}}(t,t_0) = \id_\Hil
\qquad
\ope{U}^{\mathrm{S}\to\mathrm{H}}(t,t_0)
	= (\ope{U}^{\mathrm{S}}(t,t_0))^{-1} = \ope{U}^{\mathrm{S}}(t_0,t)
	\end{split}
\\[0.5ex]			\label{12.90b}
	\begin{split}
\underline{\ope{H}}^{\mathrm{I}}(t) = \ope[\prime]{H}^{\mathrm{I}}(t)
\quad
\ope{U}^{\mathrm{I}}(t,t_0)
	= \ope[0]{U}^{\mathrm{S}}(t_0,t) \circ \ope{U}^{\mathrm{S}}(t_0,t)
\quad
\ope{U}^{\mathrm{S}\to\mathrm{I}}(t,t_0)
	= (\ope[0]{U}^{\mathrm{S}}(t,t_0))^{-1}
	= \ope[0]{U}^{\mathrm{S}}(t_0,t)
	\end{split}
	\end{align}
	\end{subequations}
where
	\begin{equation}	\label{12.91}
\ope[\prime]{H}^{\mathrm{I}}(t)
:=
\ope[0]{U}^{\mathrm{S}}(t_0,t)
\circ \ope[\prime]{H}^{\mathrm{S}}(t) \circ
\ope[0]{U}^{\mathrm{S}}(t,t_0)
\qquad
\ope[\prime]{H}^{\mathrm{S}}(t)
:=
\int \ope[\prime]{H}^{\mathrm{S}}(x_t) \Id^3\bs x
	\end{equation}
with
\(
\ope{H}^{\mathrm{S}}(t)
=
\ope[0]{H}^{\mathrm{S}}(t) + \ope[\prime]{H}^{\mathrm{S}}(t) ,
\)
$ \ope[\prime]{H}^{\mathrm{S}}(t)$ being the interaction Hamiltonian in
Schr\"odinger picture and $\ope[0]{U}^{\mathrm{S}}(t,t_0)$ being the `free'
evolution operator in Schr\"odinger picture,
	\begin{align}	\label{12.92}
\frac{\pd\ope[0]{U}^{\mathrm{S}}(t,t_0)}{\pd t}
=
\ope[0]{H}^{\mathrm{S}}(t)\circ \ope[0]{U}^{\mathrm{S}}(t,t_0)
\qquad
\ope[0]{U}^{\mathrm{S}}(t_0,t_0)
=
\id_\Hil .
	\end{align}
Besides, in accord with~\eref{12.85new}, the operator
	\begin{equation}\tag{\ref{12.90}c}	\label{12.90c}
\ope{U}^{\mathrm{H}\to\mathrm{I}}(t,t_0)
=
\ope[0]{U}^{\mathrm{S}}(t,t_0) \circ \ope{U}^{\mathrm{S}}(t,t_0)
=
\ope{U}^{\mathrm{I}}(t,t_0)
	\end{equation}
is responsible for the transition
$\omega=\mathrm{H}\mapsto\omega'=\mathrm{I}$.

	If the equations of motion for the field operators are known in a
picture $\omega$, in other picture $\omega'$ they can be derive from the
equality
	\begin{multline}	\label{12.92new}
\ih \frac{\pd\ope{A}^{\omega'}(x_t)}{\pd t}
=
	\ope{U}^{\omega\to\omega'}(t,t_0)
\circ \ih \frac{\pd\ope{A}^{\omega}(x_t)}{\pd t} \circ
	(\ope{U}^{\omega\to\omega'}(t,t_0) )^{-1}
\\
+
[
\underline{\ope{H}}^{\omega'}(t)
	- \underline{\ope{H}}^{\omega\to\omega'}(t)
,
\ope{A}^{\omega'}(x_t)
]_{\_}
	\end{multline}
which is a consequence of~\eref{12.86b}. Here
	\begin{equation}	\label{12.92new1}
\underline{\ope{H}}^{\omega\to\omega'}(t)
:=
\ope{U}^{\omega\to\omega'}(t,t_0)
\circ\ope{H}^\omega(t) \circ
(\ope{U}^{\omega\to\omega'}(t,t_0))^{-1} .
	\end{equation}
Since~\eref{12.92new} can be obtained from~\eref{12.63} by the changes
	\begin{equation}	\label{12.93}
	\begin{split}
\ope{A}(x_t) & \mapsto \ope{A}^{\omega'}(x_t)
\qquad\qquad\text{\hphantom{$_i$}}
\tope{A}(x_t)\mapsto \ope{A}^{\omega}(x_t)
\\
\ope{U}(t,t_0) & \mapsto \ope{U}^{\omega\to\omega'}(t,t_0)
\qquad\text{\hphantom{$x$}}
\ope[\prime]{H}(t)\mapsto
	  \underline{\ope{H}}^{\omega'}(t)
	- \underline{\ope{H}}^{\omega\to\omega'}(t) ,
	\end{split}
	\end{equation}
we can immediately obtain, from~\eref{12.68}, the Euler-Lagrange equation of
motion in a picture $\omega'$ if in a picture $\omega$ they are~\eref{4.1}
(with $\ope{L}^\omega$ for $\ope{L}$ and $\varphi^\omega_i(x_t)$ for
$\tope{\varphi}_i(x_t)$):

	\begin{equation}	\label{12.94}
\frac{\pd\ope{L}^{\omega'}}{\pd \ope{\varphi}_i^{\omega'}(x_t)}
-
\frac{\pd}{\pd x^\mu}\Bigl(
	\frac{\pd\ope{L}^{\omega'}}{\pd(\pd_\mu\ope{\varphi}_i^{\omega'}(x_t) )}
	 \Bigr)
=
- \frac{1}{\ih}
\Bigl[
\underline{\ope{H}}^{\omega'}(t) - \underline{\ope{H}}^{\omega\to\omega'}(t)
,
\frac{\pd\ope{L}^{\omega'}} {\pd(\pd_t\ope{\varphi}_i^{\omega'}(x_t) )}
\Bigr]_{\_} .
	\end{equation}
The same equations, of course, can be derived if one follows step-by-step the
procedure for derivation of~\eref{12.68} in which~\eref{12.66} will be
replaced by
	\begin{equation}	\label{12.95}
\frac{1}{\ih}
\frac{\pd\ope{L}^{\omega'}}{\pd(\pd_t\ope{\varphi}_j^{\omega'}(x_t) )} \circ
\Bigl[
\frac{\pd (\underline{\ope{H}}^{\omega'}(t)
	   - \underline{\ope{H}}^{\omega\to\omega'}(t) )
     }
     {\pd(\pd_t\tope{\varphi}_i^{\omega}(x_t) )}
,
\ope{\varphi}_j^{\omega'}(x_t)
\Bigr]_{\_}
= 0 .
	\end{equation}

	Similarly (see~\eref{12.92new}), if in a picture $\omega$ the
Heisenberg equations of motion are
	\begin{equation}	\label{12.96}
\ih \frac{\pd\varphi_i^\omega(x_t)}{\pd x^\mu}
=
[ \varphi_i^\omega(x_t) ,
\ope[\omega]{P}_{\mu} ]_{\_} ,
	\end{equation}
then in a picture $\omega'$ they are (cf.~\eref{12.71})
	\begin{subequations}	\label{12.97}
	\begin{align}		\label{12.97a}
\ih\frac{\pd}{\pd t} \varphi_i^{\omega'}(x_t)
& =
[
\varphi_i^{\omega'}(x_t),
	c( \ope[\omega']{P}_0 )
	- \underline{\ope{H}}^{\omega'}(t)
	+ \underline{\ope{H}}^{\omega\to\omega'}(t)
 ]_{\_}
\\				\label{12.97b}
\ih\frac{\pd}{\pd x^a} \varphi_i(x_t)
& =
[\varphi_i(x_t), \ope{P}_a ]_{\_}
\qquad
a=1,2,3 .
	\end{align}
	\end{subequations}

	If $\omega=\mathrm{H}$ and $\omega'=\mathrm{I}$, in view
of~\eref{12.90}, it is a simple checking to be seem that the above results
reduce to the corresponding ones from Sect.~\ref{Sect4}.

	At this point, we would like to show how the analogue
of the `general' picture of motion from quantum
mechanics~\cite[subsec.~2.3]{bp-BQM-pictures+integrals} can be described via
the scheme presented here. Suppose, in some  $\omega$\ndash picture of motion
are known the time equations of motion for the state vectors,
i.e.~\eref{12.85}, and for the observables, \ie
	\begin{equation}	\label{12.98}
\ih \frac{\pd\ope{A}^{\omega}(x_t)}{\pd t}
=
F( \ope{A}^{\omega}, \underline{\ope{H}}^{\omega},\ldots )
	\end{equation}
for some operator-valued function $F$ which is polynomial or convergent power
series in its operator arguments. The problem is, if there is given a
\emph{unitary} operator $\ope{V}(t_1,t)\colon\Hil\to\Hil$, to describe the
quantum evolution (in time) in the picture $\omega'=\ope{V}$ with
`$\ope{U}$\ndash operator'
	\begin{equation}	\label{12.99}
\ope{U}^{\omega\to\ope{V}} (t,t_1) = \ope{V} (t_1,t) .
	\end{equation}
	From~\eref{12.85} and~\eref{12.86a}, we derive the (time) equations
of motion for state vectors in the $\ope{V}$\ndash picture:
	\begin{align}	\label{12.100}
\frac{\pd\ope{X}_{t_1}^{\ope{V}}(x_t)}{\pd t}
=
\underline{\ope{H}}_{t_1}^{\ope{V}}(t)
			\bigl(\ope{X}_{t_1}^{\ope{V}}(x_t)\bigr)
	\end{align}
where
	\begin{align}  	\label{12.101}
& \underline{\ope{H}}_{t_1}^{\ope{V}}(t)
 :=
\ope{V}(t_1,t)
 	\circ \overline{\underline{\ope{H}}} _{t_1}^{\ope{V}}(t) \circ
 \ope{V}^{-1}(t_1,t)
=
 \underline{\ope{H}}_{t_1}^{\omega\to\ope{V}} (t)
	- \lindex[{\ope{H}}] {\ope{V}}{} _{t_1}^{\ope{V}} (t)
\\			\label{12.102}
& \overline{\underline{\ope{H}}} _{t_1}^{\ope{V}}(t)
 :=
\underline{\ope{H}}^\omega(t)
- \lindex[\underline{\ope{H}}] {\ope{V}}{} (t_1,t)
\qquad
\lindex[\underline{\ope{H}}] {\ope{V}}{} (t_1,t)
=
\ih \frac{\pd\ope{V}^{-1}(t_1,t)}{\pd t} \circ \ope{V}(t_1,t)
\\			\label{12.103}
& \underline{\ope{H}}_{t_1}^{\omega\to\ope{V}} (t)
=
	\ope{V}(t_1,t)
\circ \underline{\ope{H}}^{\omega} (t) \circ
	\ope{V}^{-1}(t_1,t)
\\ 			\label{12.103new}
& \lindex[\ope{H}] {\ope{V}}{} _{t_1}^{\ope{V}} (t)
=
	\ope{V}(t_1,t)
\circ \lindex[\ope{H}]{\ope{V}}{} (t_1,t) \circ
	\ope{V}^{-1}(t_1,t) .
	\end{align}
Respectively, the (time) evolution operator $\ope{U}^{\ope{V}}(t,t_1,t_0)$,
corresponding to equation~\eref{12.100}, is the solution of the initial\ndash
value problem
	\begin{align}	\label{12.104}
\ih \frac{\pd \ope{U}^{\ope{V}}(t,t_1,t_0) }{\partial t}
=
\underline{\ope{H}}_{t_1}^{\ope{V}} (t) \circ\ope{U}^{\ope{V}}(t,t_1,t_0)
\qquad
\ope{U}^{\ope{V}}(t_0,t_1,t_0) = \id_\Hil
	\end{align}
and determines the evolution of state vectors, \ie
	\begin{align}	\label{12.105}
\ope{X}_{t_1}^{\ope{V}}(x_t)
=
\ope{U}^{\ope{V}}(t,t_1,t_0)  ( \ope{X}_{t_1}^{\ope{V}}(x_{t_0}) ) .
	\end{align}

	The Heisenberg equations of motion for the observables in the
$\ope{V}$\ndash picture follow from~\eref{12.98} and~\eref{12.92new} and are:
	\begin{equation}	\label{12.106}
\ih \frac{\pd\ope{A}^{\ope{V}}(x_t)}{\pd t}
=
F( \ope{A}^{\ope{V}}, \underline{\ope{H}}^{\omega\to\ope{V}},\ldots )
+
[
\ope{A}^{\ope{V}}(x_t)
,
 \lindex[\underline{\ope{H}}]{\ope{V}}{} _{t_1}^{\ope{V}}(t)
]_{\_} ,
	\end{equation}
where, in view of~\eref{12.101},
\(
\underline{\ope{H}}^{\omega'}(t) - \underline{\ope{H}}^{\omega\to\omega'}(t)
=
\underline{\ope{H}}^{\ope{V}}(t) - \underline{\ope{H}}^{\omega\to\ope{V}}(t)
=\lindex[\ope{H}]{\ope{V}}{} ^{\ope{V}} ,
\)
 $\omega'=\ope{V}$, was used.

	If $\omega=\mathrm{S}$, \ie if the Schr\"odinger picture is taken as a
basic one to start off, the above results reproduce part of the ones
in~\cite[subsec.~2.3]{bp-BQM-pictures+integrals}.

	We leave the derivation of the Euler-Lagrange equation of motion in
the  $\ope{V}$\ndash picture to the reader, as an exercise.

	The time-dependent pictures considered above are
\emph{explicitly time\ndash dependent} but they are \emph{implicitly
covariant}. This can be proved by replacing the state vectors and field (and
other) operators by corresponding state functionals and field (and other
operator) functionals  depending on a space\ndash like surface $\sigma$ from
a foliation $\Sigma$ of the spacetime $\base$. The basic moment is the
replacement of~\eref{12.83} and~\eref{12.84} by, respectively,
Tomonaga\ndash Schwinger equation (cf~\eref{12.16})
	\begin{equation}	\label{12.106new1}
\ih c \frac{\delta\ope{U}^\omega[\sigma,\sigma_0]}{\delta\sigma(x)}
=
\ope{H}^\omega[\sigma;x]\circ \ope{U}^\omega[\sigma,\sigma_0]
\quad
x\in\sigma
	\end{equation}
and the initial condition (cf.~\eref{12.17})
	\begin{equation}	\label{12.106new2}
 \ope{U}^\omega[\sigma_0,\sigma_0] = \id_\Hil.
	\end{equation}
Here
	\begin{equation}	\label{12.106new3}
\ope{H}^\omega[\sigma;x]
:=
\ope{U}^\omega[\sigma,\sigma_0]
\circ \ope{H}^\omega(x) \circ
(\ope{U}^\omega[\sigma,\sigma_0])^{-1}
	\end{equation}
with $\ope{H}^\omega(x)$ being the Hamiltonian in time-dependent
$\omega$\ndash picture. (Note the connection
 $\underline{\ope{H}}^\omega(t)=\int\ope{H}^\omega(x_t)\Id^3\bs x$
with the Hamiltonian appearing in~\eref{12.83}.) Then, the
transformations~\eref{12.18}, with the just defined
`$\ope{U}$\ndash operator' $\ope{U}^\omega[\sigma,\sigma_0]$, realize the
transition to the new \emph{covariant} $\omega$\ndash picture. Further one
should follow the exposition of the covariant interaction picture of
Sect.~\ref{Sect3} with the only change that the interaction Hamiltonian(s)
must be replace by the total Hamiltonian(s), \viz
$\ope[\prime]{H}[\sigma;x]\mapsto\ope{H}^\omega[\sigma;x]$ etc. The concrete
results of the realization of this procedure are similar to the ones of
Sect.~\ref{Sect3} and at the end of Sect.~\ref{Sect5} and
will not be written here.


\section{The momentum picture}
	\label{Sect7}

	Regardless of the existence of a covariant formulation of the well
known standard Schr\"odinger picture, it still has tracks of a time\ndash
dependence: as the surfaces $\sigma$ and $\sigma_0$, as well as other ones
`between' them, must belong to a family of surfaces, forming a foliation
$\Sigma$ of the spacetime $\base$. These surfaces should be `labeled' somehow
which, in a sense, is equivalent to to the (implicit) introduction of a time
coordinate.%
\footnote{~%
The same arguments are valid for any time-dependent picture of motion, not
only for the Schr\"odinger one.%
}
Our opinion on this phenomenon is that the Schr\"odinger picture,
as considered in the literature and in Sect.~\ref{Sect5}, does not
correspond to the (special relativistic) spirit of quantum field theory and
simply partially copies a similar situation in quantum mechanics. Indeed, in
quantum mechanics, in the Schr\"odinger/Heisenberg picture, the wavefunctions
depend/don't depend on the time, while for the observables the situation is an
opposite one. This state of affairs is (\emph{mutatis mutandis}) transferred
in quantum field theory. But in it there are four, not one as in quantum
mechanics, coordinates. As a result in the Heisenberg picture, as everywhere
is accepted, the state vectors are \emph{constant in spacetime}, not only in
time, while the observables and field operators depend on a spacetime point
where they are evaluated. Correspondingly, by our opinion, it is quite more
natural one to expect that in the Schr\"odinger picture of motion of quantum
field theory the state vectors to depend on a spacetime point, not only on
its time coordinate, and the field operators (and observables constructed
from them) to be constant in spacetime, not only in time. Such a picture
(representation) of (canonical) quantum field theory  exists and we call it
the \emph{momentum picture}.%
\footnote{~%
The term 4-dimensional momentum picture is also suitable because, as we
shall see below, there exist `intermediate' momentum pictures in which
the state vectors depend on some $k$, $k=0,1,2,3$, of the coordinates, while
the field operators do not depend on these $k$ coordinates; the cases $k=0$
and $k=1$ reproduce the standard Heisenberg and Schr\"odinger pictures,
respectively, and the case $k=4$ gives the momentum picture. Elsewhere we
shall show that in momentum picture are reproduced all results from the
momentum representation of Heisenberg picture of motion. The name `momentum
picture' comes from here and the essence of the proposed new picture of
motion.%
}
Below we describe the basic characteristics of this new picture of quantum
field theory.

	Let $x,x_0\in\base$ and $\tope{P}_{\mu}$ be the
4-momentum operator of a system of quantum fields. By $\lambda\in\{0,1,2,3\}$
we shall denote a spacetime index over which a summation is \emph{not assumed}
when it appearance more than ones in some expression.

	Define `$\ope{U}$-operators'
$\ope{U}_\lambda(x^\lambda,x_0^\lambda)\colon\Hil\to\Hil$, $\lambda=0,1,2,3$
as solutions of the initial\ndash value problems
	\begin{gather}	\label{12.107}
\ih \frac{\pd \ope{U}_\lambda(x^\lambda,x_0^\lambda)}{\pd x^\lambda}
=
\ope{P}_{\lambda}(x) \circ \ope{U}_\lambda(x^\lambda,x_0^\lambda)
	\end{gather}
where%
\footnote{~%
The last equality in~\eref{12.108} follows from the commutativity between
$\ope{U}_\lambda(x^\lambda,x_0^\lambda)$ and $\tope{P}_{\mu}$ or
$\ope{P}_{\mu}$ as the last two operators are constant.%
}%
	\begin{gather}	\label{12.108}
\ope{P}_{\lambda}(x)
:=
\ope{U}_\lambda(x^\lambda,x_0^\lambda) \circ
\tope{P}_{\lambda}
\circ \ope{U}_\lambda^{-1}(x^\lambda,x_0^\lambda).
	\end{gather}
Notice, since $\tope{P}_0=\frac{1}{c}\tope{H}$ and $x^0=ct$, the
operator $\ope{U}_0(x^0,x_0^0)\equiv\ope{U}(t,t_0)$ is the same
one, defined via~\eref{12.73} and~\eref{12.57}, by means of which the
transition form Heisenberg to (time\ndash dependent) Schr\"odinger picture is
performed. Since $\tope{P}_{\mu}$  are constant, independent of
$x$ or $x_0$, operators, which expresses the energy\ndash momentum
conservation for translation invariant systems, the explicit form of
 $\ope{U}_\lambda(x^\lambda,x_0^\lambda)$ is
	\begin{equation}	\label{12.109}
\ope{U}_\lambda(x^\lambda,x_0^\lambda)
=
\e^{\iih(x^\lambda-x_0^\lambda)\tope{P}_{\lambda}}
=
\e^{\iih(x^\lambda-x_0^\lambda)\ope{P}_{\lambda}}
	\end{equation}
where $\lambda$ is \emph{not} a summation index and~\eref{12.108} was applied.

	The operator~\eref{12.109}, as well as the operators%
\footnote{~%
Since the components $\ope{P}_\mu$ of the momentum operator
commute~\cite{bp-QFT-momentum-operator} and  $(x^\lambda-x_0^\lambda)$ is
considered as a real parameter by which $\ope{P}_\lambda$ is multiplied,
the operators $\ope{U}_\lambda(x^\lambda,x_0^\lambda)$ also commute, \ie we
have
\(
[
 \ope{U}_{\lambda_1}(x^{\lambda_1},x_0^{\lambda_1})
,
\ope{U}_{\lambda_2}(x^{\lambda_2},x_0^{\lambda_2})
]_{\_} = 0 .
\)
Therefore the order in which the mappings
$\ope{U}_\lambda(x^\lambda,x_0^\lambda)$ appear
in~\eref{12.110}--\eref{12.112} below is inessential.%
}
	\begin{equation}	\label{12.110}
\ope{U}_{\lambda_1,\lambda_2}
	(x^{\lambda_1},x^{\lambda_2},x_0^{\lambda_1},x_0^{\lambda_2})
:=
\ope{U}_{\lambda_1}(x^{\lambda_1},x_0^{\lambda_1})\circ
\ope{U}_{\lambda_2}(x^{\lambda_2},x_0^{\lambda_2})
=
\exp\Bigl(\iih \sum_{\lambda=\lambda_1,\lambda_2}
		(x^\lambda-x_0^\lambda)\ope{P}_{\lambda}
    \Bigr)
	\end{equation}
\vspace{-3ex}
	\begin{multline}	\label{12.111}
\ope{U}_{\lambda_1,\lambda_2,\lambda_3}
	(x^{\lambda_1},x^{\lambda_2},x^{\lambda_3},
			x_0^{\lambda_1},x_0^{\lambda_2},x_0^{\lambda_3})
:=
\ope{U}_{\lambda_1}(x^{\lambda_1},x_0^{\lambda_1})\circ
\ope{U}_{\lambda_2}(x^{\lambda_2},x_0^{\lambda_2})\circ
\ope{U}_{\lambda_3}(x^{\lambda_3},x_0^{\lambda_3})
\\ =
\exp\Bigl(\iih \sum_{\lambda=\lambda_1,\lambda_2,\lambda_3}
		(x^\lambda-x_0^\lambda)\ope{P}_{\lambda}
    \Bigr)
	\end{multline}
\vspace{-3ex}
	\begin{multline}	\label{12.112}
\ope{U}(x,x_0)
:=
\ope{U}_{0}(x^{0},x_0^{0})\circ
\ope{U}_{1}(x^{1},x_0^{1})\circ
\ope{U}_{2}(x^{2},x_0^{2})\circ
\ope{U}_{3}(x^{3},x_0^{3})
\\ =
\exp\Bigl( \iih \sum_\mu (x^\mu-x_0^\mu)\ope{P}_{\mu} \Bigr) ,
	\end{multline}
where $\lambda_1,\lambda_2,\lambda_3\in\{0,1,2,3\}$ are \emph{different} and
$\mu$ is ordinary summation index, can be taken as `$\ope{U}$\ndash
operators' and via the transformations~\eref{12.14} and~\eref{12.15} define a
transition to new pictures of motion, which we call
$k$\ndash\emph{dimensional momentum pictures} with k=1,2,3,4
for~\eref{12.109}--\eref{12.112} respectively.  Evidently, the case $k=1$
corresponds to the ordinary, time\ndash dependent, Schr\"odinger picture. For
completeness, the case $k=0$ will be identified with the Heisenberg picture
of motion. Below we shall be interested in the case $k=4$ for which the
special name the \emph{momentum picture} will be used.

	In a sense of~\eref{12.112} (and its consequences presented below),
this new picture of motion is a composition (product) of four ordinary
(coordinate dependent) Schr\"odinger pictures, by one for each of the four
spacetime coordinates. As a result of this, one can expect in the momentum
picture the field operators (and observables which are polynomial in them) to
be constant in spacetime, contrary to the state vectors. Such a conclusion is
immediately confirmed by the observation that~\eref{12.112} is exactly the
(representation of the) spacetime translation operator (with parameter
$-(x-x_0)=x_0-x$) acting on the operators on system's Hilbert space.

	By~\eref{12.14}, the transition from Heisenberg to momentum picture is
given by the formulae:
	\begin{align}	\label{12.113}
\tope{X}\mapsto \ope{X}(x)
	&= \ope{U}(x,x_0) (\tope{X})
\\			\label{12.114}
\tope{A}(x)\mapsto \ope{A}(x)
	&= \ope{U}(x,x_0)\circ \tope{A}(x) \circ \ope{U}^{-1}(x,x_0) .
	\end{align}
In particular, the field operators transform as
	\begin{align}	\label{12.115}
\tope{\varphi}_i(x)\mapsto \ope{\varphi}_i(x)
     = \ope{U}(x,x_0)\circ \tope{\varphi}_i(x) \circ \ope{U}^{-1}(x,x_0) .
	\end{align}

	Since from~\eref{12.112} and~\eref{12.107} follows
	\begin{equation}	\label{12.116}
\ih \frac{\pd\ope{U}(x,x_0)}{\pd x^\mu}
=
\ope{P}_{\mu}\circ \ope{U}(x,x_0)
\qquad
\ope{U}(x_0,x_0) = \id_\Hil ,
	\end{equation}
due to~\eref{12.113}, we see that the state vectors $\ope{X}(x)$ in
momentum picture are solutions of the initial\ndash value problem
	\begin{equation}	\label{12.117}
\ih \frac{\pd\ope{X}(x)}{\pd x^\mu}
=
\ope{P}_{\mu} (\ope{X}(x))
\qquad
\ope{X}(x)|_{x=x_0}=\ope{X}(x_0) = \tope{X}
	\end{equation}
which is the 4-dimensional analogue of a similar problem for the Schr\"odinger
equation in quantum mechanics.

	By virtue of~\eref{12.112}, or in view of the independence of
$\ope{P}_{\mu}$ of $x$, the solution of~\eref{12.117} is
	\begin{equation}	\label{12.118}
\ope{X}(x)
=
\e^{\iih(x^\mu-x_0^\mu)\ope{P}_{\mu}} (\ope{X}(x_0)).
	\end{equation}
Thus, if $\ope{X}(x_0)=\tope{X}$  is an eigenvector of
$\ope{P}_{\mu}$ ($=\tope{P}_{\mu}$)
with eigenvalues $p_\mu$,
	\begin{equation}	\label{12.119}
\ope{P}_{\mu} (\ope{X}(x_0)) = p_\mu \ope{X}(x_0)
\quad
( =p_\mu \tope{X} = \tope{P}_{\mu} (\tope{X}) ) ,
	\end{equation}
we have the following \emph{explicit} form of the state vectors
	\begin{equation}	\label{12.120}
\ope{X}(x)
=
\e^{ \iih(x^\mu-x_0^\mu)p_\mu } (\ope{X}(x_0)).
	\end{equation}
It should clearly be understood, \emph{this is the general form of all state
vectors} as they are eigenvectors of all (commuting)
observables~\cite[p.~59]{Roman-QFT}, in particular, of the 4\ndash momentum
operator.

	To derive the equations of motion for the field operators and
observables in the new momentum picture, we shall apply the equality
	\begin{align}	\label{12.121}
\ih \frac{\pd\ope{A}(x)} {\pd x^\mu}
& =
\ope{U}(x,x_0)\circ
\ih \frac{\pd\tope{A}(x)} {\pd x^\mu}
\circ \ope{U}^{-1}(x,x_0)
+
[ \ope{P}_{\mu} , \ope{A}(x) ]_{\_}
\\ \intertext{which is a consequence of~\eref{12.114} and~\eref{12.116}. The
particular choice $\ope{A}=\varphi_i$ results in}
 			 \label{12.121new}
\ih \frac{\pd\varphi_i(x)} {\pd x^\mu}
& =
\ope{U}(x,x_0)\circ
\ih \frac{\pd\tope{\varphi}_i(x)} {\pd x^\mu}
\circ \ope{U}^{-1}(x,x_0)
+
[ \ope{P}_{\mu} , \varphi_i(x) ]_{\_}
	\end{align}
due to~\eref{12.115}. Substituting here the Heisenberg equations of
motion~\eref{6.8}, we get
	\begin{align}	\label{12.122}
\frac{\pd\varphi_i(x)}{\pd x^\mu} = 0
\intertext{or}
			\label{12.123}
\varphi_i(x) = \varphi_i(x_0)
	\end{align}
which means that \emph{in momentum picture the field operators are constant
operators}, \ie they are spacetime\ndash independent operators. This is also
an evident corollary of the fact that the induced by the
operator~\eref{12.112} action on operators simply translates their arguments
by the value $-(x-x_0)=x_0-x$. Evidently, a similar result,
	\begin{equation}	\label{12.124}
\ope{A}(x) = \ope{A}(x_0),
	\end{equation}
is valid for any observable constructed from the field operators and their
partial derivatives of finite order (in Heisenberg picture) as a polynomial
or convergent power series.

	A natural question now arises: What happens with the Euler-Lagrange
equations for the field operators in momentum picture? The answer turns to be
quite amazing and natural at the same time: they transform into an
\emph{algebraic} equations for the constant field operators~\eref{12.123}.

	Performing a calculation similar to the one in~\eref{12.67}, we find
the functional form of the Lagrangian in momentum picture
(see equations~\eref{12.121new} and~\eref{12.122})
	\begin{align}	\notag
\ope{L}
: & = \ope{L}(\varphi_i(x))
:=
\ope{U}(x,x_0)\circ
\tope{L}( \tope{\varphi}_i(x),\pd_\mu\tope{\varphi}_j(x) )
\circ \ope{U}^{-1}(x,x_0)
\\ \notag & =
\tope{L} \bigl(
\ope{U}(x,x_0)\circ  \tope{\varphi}_i(x)  \circ \ope{U}^{-1}(x,x_0)
,
\ope{U}(x,x_0)\circ  \pd_\mu\tope{\varphi}_j(x)  \circ \ope{U}^{-1}(x,x_0)
\bigr)
\\ 
	\label{12.125}
& =
\tope{L} \bigl( \ope{\varphi}_i(x)
,
\pd_\mu\ope{\varphi}_j(x)
- \iih [\ope{P}_\mu,\varphi_j(x) ]_{\_}
\bigr)
=
\tope{L} \bigl( \ope{\varphi}_i
,
- \iih [\ope{P}_\mu,\varphi_j ]_{\_}
\bigr)
	\end{align}
where we have omit the argument $x$ in the last row as all quantities in
it are constant in spacetime. Now we shall transform the Euler-Lagrange
equations~\eref{4.1} into momentum picture. The first term in~\eref{4.1}
transforms into
	\begin{equation*}
\ope{U}(x,x_0)\circ
	\frac{\pd\tope{L}} {\pd\tope{\varphi}_i(x)}
\circ \ope{U}^{-1}(x,x_0)
=
\frac{\pd\ope{L}} {\pd\tope{\varphi}_i(x)}
 =
\frac{\pd\tope{L}(\varphi_j(x),y_{j\mu})} {\pd\tope{\varphi}_i(x)}
=
\frac{\pd\tope{L}(\varphi_j(x),y_{j\mu})} {\pd\ope{\varphi}_i(x)}
=
\frac{\pd\ope{L}} {\pd\ope{\varphi}_i(x)}
	\end{equation*}
where
	\begin{equation}	\label{12.126new}
y_{j\mu}
:= - \iih [\ope{P}_\mu,\varphi_j]_{\_}
 = + \iih [\varphi_j,\ope{P}_\mu]_{\_}
	\end{equation}
and we have used that
$ \frac{\pd}{\pd\tope{\varphi}_i(x)} = \frac{\pd}{\pd\ope{\varphi}(x)_i} $
as a result of~\eref{12.115},
$ \frac{\pd\ope{L}}{\pd y_{j\nu}} \frac{\pd y_{j\nu}}{\pd\varphi_i(x)} = 0$
as
\(
\frac{\pd y_{j\nu}}{\pd\varphi_i(x)}
=
-\iih ( \ope{P}_\nu \circ (\delta_i^j\id_\Hil)
	-  (\delta_i^j\id_\Hil) \circ \ope{P}_\nu
      )
\equiv 0,
\)
and $\frac{\pd\ope{U}(x,x_0)}{\pd\tope{\varphi}(x)}\equiv 0$. Since the
momenta
	\begin{equation}	\label{12.127}
\tope{\pi}^{i\mu} := \frac{\pd\tope{L}}{\pd(\pd_\mu\tope{\varphi}_i(x))}
	\end{equation}
conjugate to $\tope{\varphi}_i(x)$ transform into (see~\eref{12.121new} and
use that $\ope{P}_\mu$ is a functional, not a function, of
$\varphi_i$)
	\begin{multline}	\label{12.128}
\pi^{i\mu} =
\ope{U}(x,x_0)\circ  \tope{\pi}^{i\mu}  \circ \ope{U}^{-1}(x,x_0)
=
\frac{\pd\ope{L}}{\pd(\pd_\mu\tope{\varphi}_i(x))}
\\ =
\frac{\pd\tope{L}(\varphi_j,y_{j\nu})} {\pd y_{i\mu}}
=
\frac{\pd\tope{L}} {\pd(\pd_\mu\tope{\varphi}_i(x))}
	\Big|_{\pd_\nu\tope{\varphi}_j(x) =y_{j\nu}}
=
\frac{\pd\ope{L}} {\pd y_{i\mu}} ,
	\end{multline}
the second term in~\eref{4.1}, multiplied by $(-1)$, will transform into
	\begin{multline*}
\ope{U}(x,x_0)\circ \
	\frac{\pd}{\pd x^\mu} \tope{\pi}^{i\mu}
\circ \ope{U}^{-1}(x,x_0)
\\ =
\frac{\pd \ope{\pi}^{i\mu}}{\pd x^\mu}
-
\frac{\pd \ope{U}(x,x_0)}{\pd x^\mu} \circ
	\tope{\pi}^{i\mu}
\circ \ope{U}^{-1}(x,x_0)
-
\ope{U}(x,x_0) \circ
	\tope{\pi}^{i\mu}
\circ \frac{\pd \ope{U}^{-1}(x,x_0)}{\pd x^\mu}
\\ =
\frac{\pd \ope{\pi}^{i\mu}}{\pd x^\mu}
-
\iih [ \ope{P}_{\mu} , \ope{\pi}^{i\mu} ]_{\_}
=
\iih [\ope{\pi}^{i\mu} , \ope{P}_{\mu} ]_{\_}
	\end{multline*}
where we used~\eref{12.116},
$\pd_\mu\ope{U}^{-1}=-\ope{U}^{-1}\circ(\pd_\mu\ope{U})\circ\ope{U}^{-1}$,
and $\pd_\nu\ope{\pi}^{i\mu}\equiv 0$ by virtue of~\eref{12.122} and that
$\ope{\pi}^{i\mu}$ is polynomial or convergent power series in the field
operators (see~\eref{12.128}).

	Inserting the above results into
 $\ope{U}(x,x_0)\circ\cdots\circ\ope{U}^{-1}(x,x_0) = 0$ with the dots
denoting the l.h.s.\ of~\eref{4.1}, we, finally, get the \emph{Euler\ndash
Lagrange equations in momentum picture} as
	\begin{equation}	\label{12.129}
\Bigl\{
\frac{\pd\tope{L}(\varphi_j,y_{l\nu})} {\pd \varphi_i(x)}
-
\iih
\Bigl[
\frac{\pd\tope{L}(\varphi_j,y_{l\nu})} {\pd y_{i\mu}}
,
\ope{P}_{\mu}
\Bigr]_{\_}
\Bigr\}
\Big|_{ y_{j\nu}=\iih[\varphi_j,\ope{P}_{\nu}]_{\_} }
=0
	\end{equation}
Since $\ope{L}$ is supposed to be polynomial or convergent power series  in
its arguments, the equations~\eref{12.129}  are \emph{algebraic}, not
differential, ones (if $\ope{P}_\mu$ is considered an given known operator).
This result is a natural one in view of~\eref{12.122}.

	We shall illustrate the above general considerations on the almost
trivial example of a free Hermitian scalar field $\tope{\varphi}$, described
in Heisenberg picture by the Lagrangian
\(
\tope{L}
= - \frac{1}{2}m^2c^4\tope{\varphi}\circ\tope{\varphi}
  + c^2\hbar^2 (\pd_\mu\tope{\varphi})\circ(\pd^\mu\tope{\varphi})
=\tope{L}(\tope{\varphi},y_\nu),
\)
with $m=\const$ and  $y_\nu=\pd_\nu\tope{\varphi}$,
and satisfying the Klein\ndash Gordon equation
$(\widetilde{\square}+\frac{m^2c^2}{\hbar^2} \id_\Hil) \tope{\varphi} = 0 $,
$\widetilde{\square}:=\pd_\mu\pd^\mu$. In momentum picture
$\tope{\varphi}$ transforms into the constant operator
	\begin{equation}	\label{12.130}
\varphi(x)
= \ope{U}(x,x_0)\circ \tope{\varphi} \circ \ope{U}^{-1}(x,x_0)
= \varphi(x_0)
= \tope{\varphi}(x_0)
=: \varphi
	\end{equation}
which, in view of~\eref{12.129},
 $\frac{\pd\tope{L}} {\pd\varphi} = -m^2c^4\varphi$, and
 $\frac{\pd\tope{L}} {\pd y_\nu} = c^2\hbar^2 y_\mu\eta^{\mu\nu} $
is a solution of%
\footnote{~
As a simple exercise, the reader may wish to prove that the D'Alembert
operator (on the space of operator\ndash valued functions) in momentum
picture is
\(
\square(\cdot) =  -\frac{1}{\hbar^2}
[ [
     \cdot ,\ope{P}_\mu ]_{\_}\ope{P}_\nu
]_{\_} \eta^{\mu\nu} .
\)
(Hint: from the relation
\(
\ih\frac{\pd\ope{A}(x)}{\pd x^\mu} = [\ope{A}(x),\ope{P}_\mu^{\text{t}}]_{\_}
\)
with
$ \tope{P}_\mu^{\text{t}} = - \ih\frac{\pd}{\pd x^\mu}+ p_\mu\id_\Hil $,
 $p_\mu=\const$ (see~\cite{bp-QFT-momentum-operator}), follows that
\(
\widetilde{\square}(\tope{\varphi}_i)
=  -\frac{1}{\hbar^2}
[ [ \tope{\varphi}_i) ,
   \tope{P}_\mu^{\text{t}} ]_{\_}\tope{P}_\nu^{\text{t}} ]_{\_} \eta^{\mu\nu}
\);
now prove that in the momentum picture
$\ope{P}_\mu^{\text{t}}=\tope{P}_\mu^{\text{t}} + \ope{P}_\mu$.)
Now, the equation~\eref{12.131} follows immediately from here and the `usual'
Klein\ndash Gordon equations. Evidently, the obtained representation of the
D'Alembert operator is valid in any picture of motion.
}
	\begin{equation}	\label{12.131}
m^2c^2\varphi
-
[ [
	\varphi,\ope{P}_\mu ]_{\_},\ope{P}^\mu
]_{\_}
= 0 .
	\end{equation}
This is the \emph{Klein-Gordon equation in momentum picture}.

	The Euler-Lagrange equations~\eref{12.129} are not enough for
determination of the field operators $\varphi_i$. This is due to the simple
reason that in them enter also the components $\ope{P}_\mu$ of the
(canonical) momentum operator~\eref{4.6} which, in view
of~\eref{4.6new1}--\eref{4.6} and~\eref{2.1}, are functions (or functionals)
of the field operators. Hence, a complete system of equations for the field
operators should consists of~\eref{12.129} and an explicit connection between
them and the momentum operator.

	A detailed presentation of quantum field theory in momentum picture
will be given elsewhere.


\section{Covariant pictures of motion}
	\label{Sect8}

	Suppose $\ope{U}(x)\colon\Hil\to\Hil$ is depending on $x\in\base$
unitary operator, $\ope{U}^\dag(x)=(\ope{U}(x))^\dag=\ope{U}^{-1}(x)$.
According to the general rules~\eref{12.14} and~\eref{12.15}, the
transformations
	\begin{subequations}	\label{8.1}
	\begin{align}
			\label{8.1a}
\tope{X} \mapsto \ope{X}(x)
  & := \ope{U}(x)(\tope{X})
\\			\label{8.1b}
\tope{A}(x) \mapsto \ope{A}(x)
  & := \ope{U}(x)\circ \tope{A}(x)\circ \ope{U}^{-1}(x)
	\end{align}
	\end{subequations}
define a transition from Heisenberg picture of motion to a new one which,
for brevity, will be called the $\ope{U}$\ndash picture. Evidently, for a
fixed point $x_0\in\base$, the below written choices~\eref{8.2}
and~\eref{8.3} describe the momentum picture and the `initial' Heisenberg
one, respectively. The purpose of this section is a brief derivation of the
equations of motion in the $\ope{U}$\ndash picture.

	In accord with the general transformation~\eref{8.1b}, the partial
derivative $\pd_\mu\tope{A}(x)$ of an operator $\tope{A}(x)\colon\Hil\to\Hil$
transforms into
\(
\ope{U}(x)\circ (\pd_\mu\tope{A}(x))\circ \ope{U}^{-1}(x)
=
\pd_\mu \bigl( \ope{U}(x)\circ (\tope{A}(x))\circ \ope{U}^{-1}(x) \bigr)
- (\pd_\mu \ope{U}(x)) \circ \tope{A}(x)\circ \ope{U}^{-1}(x)
- \ope{U}(x)\circ \tope{A}(x)\circ (\pd_\mu\ope{U}^{-1}(x))
=
\pd_\mu\ope{A}(x)
	+ [\ope{A}(x), \pd_\mu \ope{U}(x) \circ \ope{U}^{-1}(x) ]_{\_} .
\)
Therefore
	\begin{gather}
			\label{8.4}
\pd_\mu \tope{A}(x)
\mapsto
\ope{U}(x)\circ (\pd_\mu\tope{A}(x))\circ \ope{U}^{-1}(x)
=
\pd_\mu\ope{A}(x)
	+ [\ope{A}(x), \ope{H}_\mu(x)]_{\_}
\\\intertext{where $\ope{A}(x)$ is given via~\eref{8.1b} and we have
introduced the shorthand}
			 \label{8.5}
\ope{H}_\mu(x) := (\pd_\mu \ope{U}(x)) \circ \ope{U}^{-1}(x) .
\\\intertext{In particular, for the field operators $\tope{\varphi}_i(x)$,
\eref{8.1b} and~\eref{8.5} respectively read}
			 \label{8.6}
\tope{\varphi}_i(x) \mapsto \ope{\varphi}_i(x)
=
\ope{U}(x)\circ \tope{\varphi}_i(x)\circ \ope{U}^{-1}(x)
\\			 \label{8.7}
\pd_\mu \tope{\varphi}_i(x)
\mapsto
\ope{U}(x)\circ (\pd_\mu\tope{\varphi}_i(x))\circ \ope{U}^{-1}(x)
=
\pd_\mu\varphi_i(x) + [\varphi_i(x), \ope{H}_\mu(x)]_{\_} .
	\end{gather}

	Let the system's Lagrangian
$\tope{L}=\tope{L}(\tope{\varphi}_i(x),\pd_\nu\tope{\varphi}_j(x))$ be
polynomial or convergent power series in the field operators and their first
partial derivatives. Then, repeating the steps in~\eref{12.125} and
applying~\eref{8.6} and~\eref{8.7}, we find the Lagrangian in the
$\ope{U}$\ndash picture as
	\begin{equation}
			\label{8.8}
\ope{L}
= \ope{L}(\varphi_i(x),\pd_\mu\varphi_j(x))
= \tope{L}(\varphi_i(x),\pd_\mu\varphi_j(x)
	+ [\varphi_j(x),\ope{H}_\mu(x)]_{\_} ) .
	\end{equation}

	Now we intend to transform the Euler-Lagrange equations~\eref{4.1} in
$\ope{U}$\ndash picture. Introducing, for brevity, the notation

	\begin{equation}
			\label{8.9}
y_{i\mu}
:=
\pd_\mu\varphi_i(x) + [\varphi_i(x), \ope{H}_\mu(x)]_{\_}
\qquad
\bigl(
= \ope{U}(x)\circ (\pd_\mu\tope{\varphi}_i(x))\circ \ope{U}^{-1}(x)
\bigr) ,
	\end{equation}
we see that the first term in the l.h.s.\ of~\eref{4.1} transforms into
\(
\ope{U}(x)\circ
	\frac{\pd\tope{L}}{\pd\tope{\varphi}_i(x)} \circ\ope{U}^{-1}(x)
=
\frac{\pd\tope{L}(\varphi_j(x),y_{j\nu})}{\pd\ope{\varphi}_i(x)} ,
\)
while the second term in it, in view of~\eref{8.4}, transforms into
\(
- \ope{U}(x) \circ \pd_\mu\tope{\pi}^{i\mu} \circ \ope{U}^{-1}(x)
=
-\pd_\mu\pi^{i\mu} - [\pi^{i\mu},\ope{H}_\mu]_{\_} ,
\)
where
$ \tope{\pi}^{i\mu} := \frac{\pd\tope{L}}{\pd_\mu\tope{\varphi}_i(x)} $
and
\(
\pi^{i\mu}
=
 \ope{U}(x) \circ \tope{\pi}^{i\mu} \circ \ope{U}^{-1}(x)
=
 \frac{\pd\ope{L}}{\pd_\mu\tope{\varphi}_i(x)}
=
 \frac{\pd\ope{L} (\varphi_j(x),y_{j\nu}) }{\pd y_{i\mu}} .
\)
Combining the above results, we get the \emph{Euler-Lagrange equations in
$\ope{U}$\ndash picture} as
	\begin{multline}	\label{8.10}
\Bigl\{
\frac{\pd\tope{L}(\varphi_j,y_{l\nu}(x))} {\pd \varphi_i(x)}
- \frac{\pd}{\pd x^\mu}
  \Bigl( \frac{\pd\tope{L}(\varphi_j(x),y_{l\nu}(x)) }{\pd y_{i\mu}(x)} \Bigr)
\\
-
\Bigl[
\frac{\pd\tope{L}(\varphi_j,y_{l\nu}(x))} {\pd y_{i\mu}(x)}
,
\ope{H}_{\mu} (x)
\Bigr]_{\_}
\Bigr\}
\Big|_{ y_{j\nu}(x)
	= \pd_\nu\varphi_j(x) + [\varphi_j(x),\ope{H}_{\nu}(x)]_{\_} }
=0 .
	\end{multline}

	In view of~\eref{6.8},~\eref{8.7} and~\eref{8.1b}, the
\emph{Heisenberg equations/relations in $\ope{U}$\ndash picture} are
	\begin{gather}	\label{8.11}
[\varphi_i(x),\ope{P}_\mu(x)]_{\_}
=
\ih \{ \pd_\mu\varphi_i(x) + [\varphi_i(x),\ope{H}_{\mu}(x)]_{\_} \}
\\\intertext{or, equivalently,}
			\label{8.12}
[\varphi_i(x),\ope{P}_\mu(x) -\ih \ope{H}_{\mu}(x)]_{\_}
=
\ih \pd_\mu\varphi_i(x)
\\\intertext{where, in conformity with~\eref{8.1b}, $\ope{P}_\mu(x)$ is the
momentum operator in  $\ope{U}$\ndash picture, \ie}
			\label{8.13}
\ope{P}_\mu(x) := \ope{U}(x)\circ \tope{P}_\mu \circ\ope{U}^{-1}(x) .
\\\intertext{Evidently, the equations~\eref{8.12} hold in any picture of
motion while~\eref{6.8} are valid if and only if}
			\label{8.14}
[\varphi_i(x),\ope{H}_{\mu}(x)]_{\_} = 0 .
	\end{gather}

	Combining~\eref{8.11} and~\eref{8.10}, we can write the
Euler-Lagrange equations in $\ope{U}$\ndash picture also as
	\begin{multline}	\label{8.15}
\Bigl\{
\frac{\pd\tope{L}(\varphi_j,y_{l\nu}(x))} {\pd \varphi_i(x)}
- \frac{\pd}{\pd x^\mu}
  \Bigl( \frac{\pd\tope{L}(\varphi_j(x),y_{l\nu}(x)) }{\pd y_{i\mu}(x)} \Bigr)
\\
-
\Bigl[
\frac{\pd\tope{L}(\varphi_j,y_{l\nu}(x))} {\pd y_{i\mu}(x)}
,
\ope{H}_{\mu}(x)
\Bigr]_{\_}
\Bigr\}
\Big|_{ y_{j\nu}(x) = \iih [\varphi_j(x),\ope{P}_{\nu}(x)]_{\_} }
= 0 .
	\end{multline}

	The spacetime evolution of the state vectors is an almost trivial
corollary of the transformation~\eref{8.1a}. Indeed, let $\ope{X}_0$ be the
value of a state vector $\ope{X}(x)$ at a fixed point $x_0\in\base$,
$\ope{X}(x_0)=\ope{X}_0$, then, by virtue of~\eref{8.1a}, we have
	\begin{gather}	\label{8.17}
\ope{X}(x) = \ope{U}(x,x_0) ( \ope{X}(x_0) )
\\\intertext{where the operator}
			\label{8.18}
\ope{U}(x,x_0) := \ope{U}(x) \circ \ope{U}^{-1}(x_0)
	\end{gather}
has a sense of an \emph{evolution operator} in $\ope{U}$-picture.

	It is obvious, the choices
	\begin{gather}
			\label{8.2}
\ope{U}(x) = \exp\Big\{\iih(x^\mu-x_0^\mu)\tope{P}_\mu\Bigr\}
\\			\label{8.3}
\ope{U}(x) = \alpha(x) \id_\Hil
\qquad
\alpha\colon\base\to\field[C]\backslash\{0\}
\\\intertext{of the operator $\ope{U}$ select the transition from
Heisenberg picture to an $\ope{U}$\ndash picture with}
			\label{8.22}
\ope{H}_\mu(x) = \iih \tope{P}_\mu(x)
\\			\label{8.23}
\ope{H}_\mu(x) = \alpha(x) (\pd_\mu\alpha(x))  \id_\Hil
	 \end{gather}
and, consequently, describe the momentum and Heisenberg pictures,
respectively.

	If one studies the angular momentum properties of a quantum system,
it may turn to be useful the `angular momentum picture', for which
	\begin{equation}	\label{8.19}
\ope{U}(x) = \e^{\iih a^{\mu\nu}(x) \ope{J}_{\mu\nu}}
	\end{equation}
where $\ope{J}_{\mu\nu}$ are the components of the (total) angular momentum
operator and $a^{\mu\nu}(x)$ are (point\ndash dependent) parameters of a
Lorentz 4\ndash rotation. Similarly, the `charge picture', for which
	\begin{equation}	\label{8.20}
\ope{U}(x) = \e^{\frac{e}{\ih c} \ope{Q}\lambda},
	\end{equation}
where $e$ is the (electric) charge constant, $\ope{Q}$ is the (total) charge
operator and $\lambda$ is a real parameter, may turn to be essential in the
study of the `charge properties' of quantum fields.

	As (non-covariant) versions of the `angular momentum picture' can be
considered the `orbital angular momentum picture' and `spin momentum picture'
for which $\ope{U}(x)$ is given by~\eref{8.19} with the orbital angular
momentum operator $\ope{L}_{\mu\nu}$ and spin angular momentum operator
$\ope{S}_{\mu\nu}$, respectively, for $\ope{J}_{\mu\nu}$.

	As (covariant) variant of the `charge picture' can be considered the
`local charge picture' or `gauge picture' described via~\eref{8.20} with
$\lambda=\lambda(x)$ being a scalar function of $x\in\base$.


\section {Conclusion}
\label{Conclusion}

	A unitary transformation of the operators and state vectors is the
leading idea of the pictures of motion in quantum field theory (and in
quantum mechanics as well). In this way, starting form some concrete
`picture', one obtains different and equivalent representations of the whole
theory. The value of these representations is that in them vanish some
quantities and in this way one can concentrate his/her attention on the
remaining ones. For instance:
\begin{description}

\item
	(i)~In Heisenberg picture vanish the partial derivatives of the
state vectors, \ie in it they are constant, contrary to the field operators.

\item
	(ii)~In Schr\"odinger picture disappear the time derivatives of the
field operators, which is balanced by appearance of time dependence in the
state vectors.

\item
	(iii)~In interaction picture are zero some terms in the Euler\ndash
Lagrange equations of motion so that they have a form of free equations.

\item
	(iv)~In momentum picture vanish the partial derivatives of the field
operators, \ie they are constant in it, but the state vectors become spacetime
dependent.

	\end{description}

	As we saw in this paper, the time-dependent pictures of motion, like
Schr\"odinger and interaction ones, admit similar `covariant' formulations,
but, by our opinion, in them the time dependence is also presented in a latent
form. The covariant pictures of motion suit best to the special relativistic
spirit of quantum field theory; typical examples of them are the Heisenberg
and momentum ones.

	In principle, one can consider also pictures of motion intermediate
between the time\ndash dependent and covariant ones, like the $k$\ndash
dimensional, $k=2,3$, momentum picture; however, it seems that at present
such pictures have no applications.

	Generally, for different purposes, different pictures of motion may
turn to be convenient.

	The proposed in this work momentum picture will be investigated at
length elsewhere.


\addcontentsline{toc}{section}{References}
\bibliography{bozhopub,bozhoref}
\bibliographystyle{unsrt}
\addcontentsline{toc}{subsubsection}{This article ends at page}

\end{document}